\documentclass[10pt, aps, prx, twocolumn, superscriptaddress, nofootinbib]{revtex4-2}

\usepackage{amsmath, amssymb, bm, mathtools, physics, array, lipsum, subcaption, graphicx, tikz, placeins}
\usepackage{bbm}
\usepackage[hidelinks]{hyperref}
\usepackage{cleveref}
\usetikzlibrary{quantikz2, arrows}

\newcolumntype{L}{>{$}l<{$}}
\newcolumntype{C}{>{$}c<{$}}
\newcolumntype{R}{>{$}r<{$}}

\begin{document}

\title{The Octo-Rail Lattice: a four-dimensional cluster state design}

\author{Emil E.B. \O stergaard}
\thanks{These authors contributed equally to this work.}
\email{eebos@dtu.dk}
\affiliation{Center for Macroscopic Quantum States (bigQ), Department of Physics, Technical University of Denmark, 2800 Kongens Lyngby, Denmark}

\author{Niklas Budinger}
\thanks{These authors contributed equally to this work.}
\email{nbudinge@t-online.de}
\affiliation{Center for Macroscopic Quantum States (bigQ), Department of Physics, Technical University of Denmark, 2800 Kongens Lyngby, Denmark}
\affiliation{Johannes-Gutenberg University of Mainz, Institute of Physics, Staudingerweg 7, 55128 Mainz, Germany}

\author{Mikkel V. Larsen}
\affiliation{Xanadu Quantum Technologies, Toronto, Canada}%

\author{Peter~van~Loock}
\affiliation{Johannes-Gutenberg University of Mainz, Institute of Physics, Staudingerweg 7, 55128 Mainz, Germany}

\author{Jonas S. Neergaard-Nielsen}
\affiliation{Center for Macroscopic Quantum States (bigQ), Department of Physics, Technical University of Denmark, 2800 Kongens Lyngby, Denmark}

\author{Ulrik L. Andersen}
\email{ulrik.andersen@fysik.dtu.dk}
\affiliation{Center for Macroscopic Quantum States (bigQ), Department of Physics, Technical University of Denmark, 2800 Kongens Lyngby, Denmark}

\date{\today}

\begin{abstract}

Macronode cluster states are a promising candidate for fault-tolerant continuous-variable quantum computing, combining gate teleportation via homodyne detection with the Gottesman-Kitaev-Preskill code for universality and error correction. While the two-dimensional Quad-Rail Lattice offers flexibility and low noise, it lacks the dimensionality required for topological error correction codes essential for fault tolerance.
This work presents a four-dimensional cluster state, termed the Octo-Rail Lattice, generated using time-domain multiplexing. This new macronode design combines the noise properties and flexibility of the Quad-Rail Lattice with the possibility to run various topological error correction codes including surface and color codes. Besides, the presented experimental setup is easily scalable and includes only static optical components allowing for a straight-forward implementation.
Analysis demonstrates that the Octo-Rail Lattice, when combined with GKP qunaught states and the surface code, exhibits noise performance compatible with a fault-tolerance threshold of 9.75 dB squeezing. Also, universality can be ensured without requiring additional resources such as other non-Gaussian states or feed-forward operations.
This finding implies that the primary challenge in constructing an optical quantum computer lies in the experimental generation of these highly non-classical states.
Finally, a generalisation of the design to arbitrary dimensions is introduced, where the setup size scales linearly with the number of dimensions. This general framework holds promise for applications such as state multiplexing and state injection.

\end{abstract}

\maketitle

\section{Introduction}\label{sec:Introduction}

Quantum computing offers the potential to efficiently solve problems considered intractable for classical computers \cite{Feynman1982-nk, Preskill2018quantumcomputingin}, promising breakthroughs across numerous domains. Applications range from simulating complex materials \cite{Bauer2020-rr, RevModPhys.86.153} and chemical reactions \cite{Cao2019-qe} to addressing optimisation problems \cite{grover1996fastquantummechanicalalgorithm, Montanaro2016-no} and breaking cryptographic protocols \cite{doi:10.1137/S0097539795293172}. Despite steady experimental progress, the practical realisation of large-scale quantum computers remains a significant challenge.

Photonic systems, particularly within measurement-based quantum computing (MBQC), provide a compelling approach to building scalable quantum computers. MBQC leverages inherent advantages of photonic platforms, such as low decoherence and room-temperature operation, to enable scalable quantum systems, replacing coherent unitary gate sequences with adaptive projective measurements on highly entangled resource states known as cluster states \cite{PhysRevLett.86.5188, PhysRevA.68.022312, PhysRevLett.97.110501}. In the continuous-variable (CV) domain, large-scale cluster states can be generated deterministically using squeezed states \cite{Larsen_2019, doi:10.1126/science.aay2645, Zhu:21}. The scalability and fault tolerance of MBQC are directly linked to the structure and quality of these cluster states, making their design a critical consideration in photonic quantum architectures.

A practical implementation for CV MBQC is using an encoding scheme that defines qubit states within the CV Hilbert space while protecting logical quantum information from noise. The Gottesman-Kitaev-Preskill (GKP) state \cite{GKP} provides a natural solution by mapping Gaussian noise in quantum operations to discrete Pauli errors, which can then be corrected using topological error correction codes, such as the surface code \cite{KitaevSurfaceCode, fowler, Litinski_2019}. Combined with easily implementable Gaussian operations \cite{PhysRevLett.123.200502, GKP, walshe_streamlined_2021}, GKP states can form the basis for universal fault-tolerant quantum computation in photonic systems. While the generation of optical GKP states with sufficiently low logical error rates remains an open experimental challenge, there has been notable recent progress. Theoretical proposals have advanced the feasibility of photonic GKP state preparation \cite{PhysRevA.110.012436}, and initial demonstrations of grid-like structures have been made by breeding cat states \cite{asavanant2024toward, konno2024logical}. Most recently, \cite{Larsen2025} reported the first direct generation of an optical GKP grid state, demonstrating significantly higher quality than previous attempts and marking an important step toward universal photonic quantum computing.

In order to incorporate GKP-encoded qubits into scalable, fault-tolerant architectures, previous work has proposed spatially distributed resource states \cite{PRXQuantum.2.040353, aghaee2025scaling}, where nearest-neighbour entanglement supports surface-code-based error correction. However, these approaches require extensive spatial multiplexing, posing practical challenges for near-term implementation.
In this work we introduce the Octo-Rail Lattice (ORL) cluster state, a four-dimensional (4D) cluster state, that offers an alternative approach by extending the entanglement structure of the Quad-Rail Lattice (QRL) into additional temporal dimensions.  For fault-tolerant quantum computation, the ORL is reduced to a three-dimensional (3D) cluster state, where two dimensions define the surface code for topological error correction, while the third dimension is used for computation. This approach preserves the same level of error correction as \cite{PRXQuantum.2.040353, aghaee2025scaling} while significantly reducing spatial resource requirements. The ORL consists entirely of passive beamsplitters and delay lines, leveraging time-multiplexed modes to construct a 3D entanglement structure within a fixed physical footprint. This approach enables near-term scalability that is limited primarily by the propagation loss accumulated in the fibre delay lines, rather than by the complexity of the underlying architecture. Furthermore, in contrast to other implementations such as photonic integrated circuits, bulk and fibre-based implementation of the ORL is immediately doable using commercially available bulk and fibre optical components with ultra low loss required for fault-tolerant operation.

We further demonstrate that using qunaught GKP input states \cite{PhysRevA.102.062411, PhysRevA.95.012305} together with heterodyne-based magic state generation \cite{PhysRevLett.123.200502} is sufficient to enable universal, fault-tolerant computation in this architecture. Beyond its realisation of the surface code, the ORL design supports generalisations to higher-dimensional topological codes and allows for efficient, switch-free state injection and mode multiplexing.

This paper is structured as follows: Section II introduces the ORL cluster state and details its generation and entanglement structure. Section III explores the use of the ORL for fault-tolerant quantum computation using the surface code, including magic state generation and universality. Section IV discusses potential extensions of the ORL design and its implications for scalable photonic quantum computing.

Throughout this paper we set $\hbar = 1$ (vacuum variance equal to $\tfrac{1}{2}$). Further conventions and notations are introduced in Appendix A. Moreover, Appendix A provides background on GKP encoding, CV cluster states including the Dual- and Quad-Rail Lattice, and topological error correction. Importantly, we graphically represent a balanced beamsplitter operation by a vertical arrow pointing from mode $j$ to mode $k$,
\begin{align*}
    \includegraphics[]{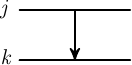}
\end{align*}

We recommend that readers not well acquainted with CV MBQC consult Appendix A.
\section{Octo-Rail Lattice Cluster State}\label{sec:ORL}

Any cluster state design utilising the error correction properties of GKP qubits necessarily needs an additional higher level qubit error correction code in order to correct logical errors and provide full fault-tolerance. When relying on nearest neighbor interactions within the cluster state, this requires an at least two-dimensional code layout alongside the one dimension used for gate implementation by teleportation. Such a design must therefore possess an at least three-dimensional connectivity. In \cite{PRXQuantum.2.040353} this is achieved by proposing a spatial, two-dimensional grid of connected QRLs.


Instead, in this work, extending the QRL design to the ORL is shown to add more temporal dimensions, resulting in scalability with a fixed number of spatial resources. The ORL cluster state is constructed using four GKP Bell pairs, depicted as thick coloured lines between two circles,
\begin{align}\label{eq:GKPBellPair2}
    \includegraphics{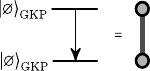}
\end{align} 
each equipped with time delays of one, $n$, $mn$ and $kmn$ clock cycles respectively, where $n$, $m$ and $k$ are non-negative integers, and connecting them by twelve beamsplitters in order to create a so-called eightsplitter. 

\begin{figure}[h]
	\centering
	\includegraphics{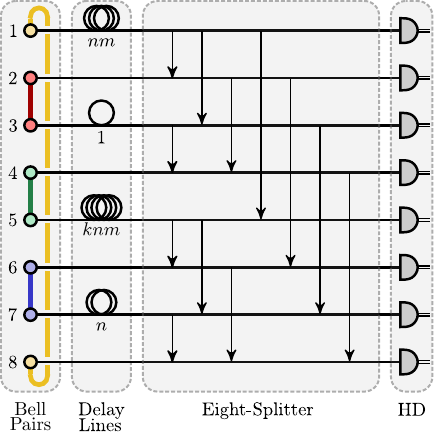}
	\caption{Setup of the Octo-Rail Lattice cluster state. Each time step, four Bell pairs are generated and partially delayed by one, $n$, $mn$ and $kmn$ clock cycles, respectively. The non-delayed halves from the current together with the delayed halves from previous time steps are then entangled by a beamsplitter network known as eightsplitter and measured by eight homodyne detectors (HD).}
	\label{fig:octo_4}
\end{figure}

\begin{figure}
	\centering
	\includegraphics[width=\columnwidth]{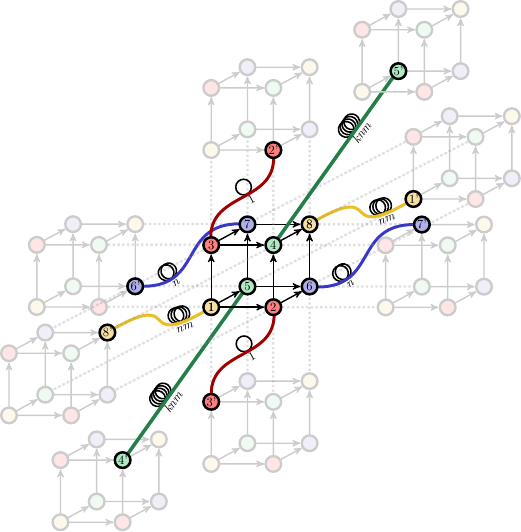}
	\caption{Depiction of a macronode of the Octo-Rail Lattice cluster state within the generated macronode lattice. The four-dimensional layout is created by the Bell pairs linking each macronode to macronodes at eight different time steps.}
	\label{fig:octo_4D_macronod}
\end{figure}

 The full setup can be seen in Fig.\,\ref{fig:octo_4} and the corresponding macronode representation shown in Fig.\,\ref{fig:octo_4D_macronod}. Consider the individual parts:
\paragraph{Bell pairs}
The system leverages the two-mode entanglement of GKP Bell pairs to interconnect different macronodes and allow for the teleportation of logical information between them. Notably, every teleportation along one of the Bell pairs inherently performs a GKP error correction on the teleported state. Four Bell pairs are required per clock cycle, and each is generated by interfering two GKP qunaught states on a beamsplitter, as shown in Eq.\,\eqref{eq:GKPBellPair2}. In \cite{PhysRevA.102.062411} it was shown, that replacing either of the two GKP qunaught states of a Bell pair by a squeezed state does not affect the performance of the teleportation, but disables the error correction in the corresponding quadrature. Given the probabilistic nature of GKP state preparation \cite{PhysRevA.110.012436}, it might prove useful to toggle between a generated qunaught state and a squeezed state depending on the success of the qunaught state generation.

\paragraph{Delay lines}
The delay lines distribute the entangled Bell pairs across macronodes and thereby define the connectivity of the macronode lattice. Choosing four delay lines which are multiples of one another, with delays of one, $n$, $mn$ and $kmn$ clock cycles, respectively, connects the macronode of clock cycle $j$ with the nodes at cycles $j+1$, $j-1$, $j+n$, $j-n$, $j+nm$, $j-nm$, $j+knm$ and $j-knm$ resulting in a four-dimensional macronode lattice with three skewed periodic boundaries. While graphical depiction becomes difficult in higher dimensions, this structure is easily captured by assigning an index $j = 0, 1, 2, ...$ to each macronode and consequently mapping it into a four-dimensional vector $(j_1, j_2, j_3, j_4)$ with,
\begin{align}
    j = j_1 + n\cdot j_2 + mn\cdot j_3 + kmn\cdot j_4.
\end{align}
In this four-dimensional lattice each macronode $(j_1, j_2, j_3, j_4)$ is connected to its eight nearest neighbors,
\begin{align}\begin{split}
    \{&\left(j_1\pm1, j_2, j_3, j_4\right),
    \left(j_1, j_2\pm1, j_3, j_4\right),\\
    &\left(j_1, j_2, j_3\pm1, j_4\right),
    \left(j_1, j_2, j_3, j_4\pm1\right)\}
\end{split}\end{align}
 by the partly delayed GKP Bell pairs.
The three skewed boundary conditions are given by,
\begin{align}\begin{split}
    \left(j_1 + n, j_2, j_3, j_4\right) = \left(j_1, j_2 + 1, j_3, j_4\right)\\
    \left(j_1, j_2 + m, j_3, j_4\right) = \left(j_1, j_2, j_3 + 1, j_4\right)\\
    \left(j_1, j_2, j_3 + k, j_4\right) = \left(j_1, j_2, j_3, j_4 + 1\right)
\end{split}\end{align}
and a unique representation of a macronode, $j$, obtained for $0\leq j_1<n$, $0\leq j_2<m$ and $0\leq j_3<k$. Hence, the size of the first three dimensions before wrapping around are decided by the three integers $n$, $m$ and $k$.

\begin{figure}
    \centering
    \begin{subfigure}[b]{\columnwidth}
        \centering
        \includegraphics{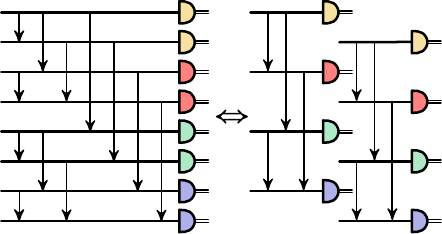}
        \caption{The DRL beamsplitter layer gets removed.}
        \label{fig:octorail_2_quadrai_1}
    \end{subfigure}
    \begin{subfigure}[b]{\columnwidth}
        \centering
        \includegraphics{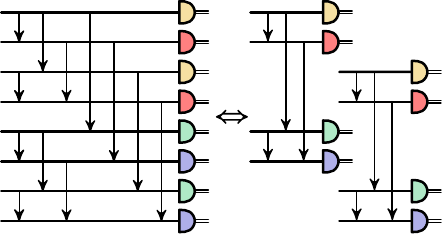}
        \caption{The QRL beamsplitter layer gets removed.}
        \label{fig:octorail_2_quadrai_2}
    \end{subfigure}
    \begin{subfigure}[b]{\columnwidth}
        \centering
        \includegraphics{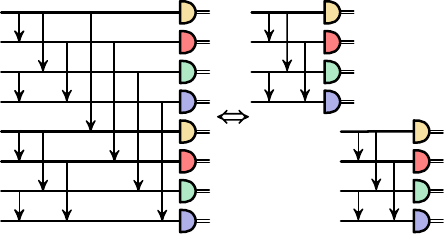}
        \caption{The ORL beamsplitter layer gets removed.}
        \label{fig:octorail_2_quadrai_3}
    \end{subfigure}
    \caption{Reduction of an ORL macronode into two separate QRL macronodes by applying the same measurement bases across modes. Equally colored detectors measure in the identical basis.}
    \label{fig:octorail_decom}
\end{figure}

\paragraph{Eightsplitter and homodyne detectors}
To enable varying gates within different macronodes, the measurement bases of the eight homodyne detectors must be dynamically adaptable. The specific gates which can be implemented as well as their gate noise levels depend on the chosen beamsplitter network. Here, the eightsplitter exhibits both high flexibility and low gate noise due to symmetries inherited from the underlying QRLs. Specifically, its three beamsplitter layers, namely the DRL, the QRL and the ORL layer, all commute, leading to,
\begin{align}
    \includegraphics[width=\columnwidth]{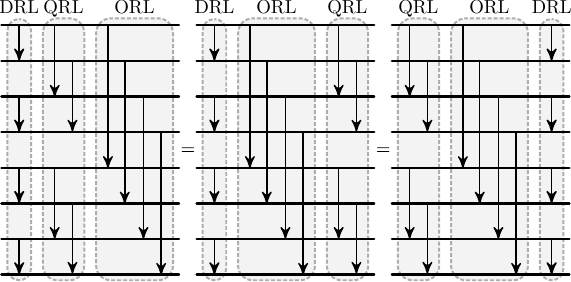}
\end{align}
By choosing identical measurement bases for appropriate pairs of modes, this structure allows the removal of any one of the three beamsplitter layers. Consequently, different pairs of QRLs can be executed on the ORL, as illustrated in Fig.\,\ref{fig:octorail_decom}. Thus the ORL inherits the respective single- and two-mode gates as well as their gate noise from the given QRLs. Additionally, a general four-mode gate $\hat{V}_4\left(\theta_1, ..., \theta_8\right)$ -- here acting on modes 1, 3, 5 and 7 -- can be expressed as,
\begin{align}\label{eq:gateV4}
\includegraphics[]{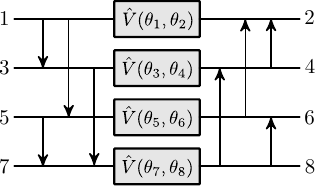}
\end{align}

\begin{table*}[ht]
    \caption{Selected measurement bases of the eight homodyne detectors of a ORL and the resulting gates.}\label{tab:ORL}
    \begin{ruledtabular}
        \begin{tabular}{CCC}
            \theta_1, \theta_2, \theta_3, \theta_4, \theta_5, \theta_6, \theta_7, \theta_8 & \hat V_4\left(\theta_1, \theta_2,  \theta_3, \theta_4, \theta_5, \theta_6, \theta_7, \theta_8\right) & \text{Logical Gate} \\ \hline \rule{0mm}{\normalbaselineskip}
            0, \frac{\pi}{2}, 0, \frac{\pi}{2}, 0, \frac{\pi}{2}, 0, \frac{\pi}{2} & \hat I \otimes \hat I \otimes \hat I \otimes \hat I & \bar I \otimes \bar I \otimes \bar I \otimes \bar I \\
            -\frac{\pi}{4}, \frac{\pi}{4}, -\frac{\pi}{4}, \frac{\pi}{4}, -\frac{\pi}{4}, \frac{\pi}{4}, -\frac{\pi}{4}, \frac{\pi}{4} & \hat F \otimes \hat F \otimes \hat F \otimes \hat F & \bar H \otimes \bar H \otimes \bar H \otimes \bar H \\
            0, -\arctan(2),  0, -\arctan(2), 0, -\arctan(2),  0, -\arctan(2) & \hat P\left(-1\right) \otimes \hat P\left(-1\right) \otimes \hat P\left(-1\right) \otimes \hat P\left(-1\right) & \bar P \otimes \bar P\otimes \bar P \otimes \bar P \\
            \frac{\pi}{2}, 0, 0, \frac{\pi}{2}, \frac{\pi}{2}, 0, 0, \frac{\pi}{2} & \text{SWAP} \otimes \text{SWAP} & \overline{\text{SWAP}} \otimes \overline{\text{SWAP}} \\
            0, -\arctan(2),  0, \arctan(2), 0, -\arctan(2),  0, \arctan(2)  & \hat C_Z(1) \otimes \hat C_Z(1) & \bar C_Z \otimes \bar C_Z
        \end{tabular}
    \end{ruledtabular}
\end{table*}

\noindent While not included here, the corresponding derivation can be obtained by closely following the derivation of the general QRL gate given in \cite{walshe_streamlined_2021}.
Some common gates for the ORL including all GKP Clifford gates along with their corresponding measurement bases are listed in Table\,\ref{tab:ORL}.

Despite being listed for only one specific arrangement of in- and output modes, the presented gates can be implemented for many more. 
To analyse this flexibility, one can treat different mode arrangements as permutations of the eight input and output modes.
Whenever commuting one of these permutations through the eightsplitter results only in a permutation of the original measurement bases alongside single-mode phase rotations, corresponding measurement bases for the new mode arrangement can easily be determined.
Out of the possible $8!=40320$ permutations, 1344 fulfill this requirement and can therefore be used to change the in- and outputs of any given gate. These allowed permutations $P_\text{allowed}\subset S_8$ form a group generated by the double transpositions,
\begin{align}
    P_\text{allowed}=\left<\big\{(12)(56),(13)(57),(14)(58),(17)(28) \right\}\big>,
\end{align}    
where the transposition $(jk)$ swaps modes $j$ and $k$. In contrast to the foursplitter, a single swap of two modes can in general not be compensated for by changing the measurement bases. In cases where this compensation is not possible, running the given gate on the desired combination of in- and output modes cannot be achieved dynamically and requires a different static configuration of the ORL. More precisely, the group $P_\text{allowed}$ decomposes $S_8$ into 30 right cosets, each corresponding to a statically different setup of the ORL able to run the same gates but on different modes. A depiction of these 30 setups can be found in the supplementary material along with an intuitive representation of $P_\text{allowed}$.
When specifically considering the single- and two-mode GKP Clifford gates, the former can be run on any combination of in- and output modes, while the latter can be run on most but not all combinations.

\section{Surface Code on the ORL}

\begin{figure}
	\centering
	\includegraphics[width=\columnwidth]{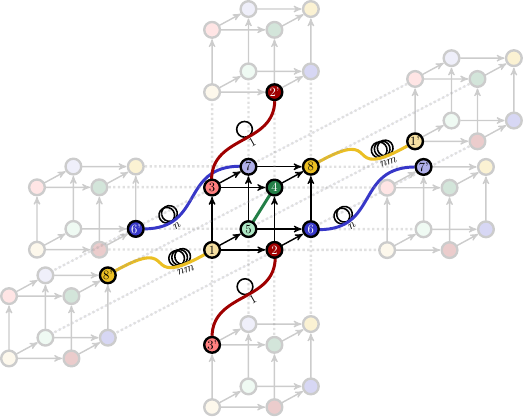}
	\caption{Depiction of a macronode of the Octo-Rail Lattice cluster state adapted to the surface code. Removing the longest delay line by setting $k=0$ reduces the macronode lattice to three dimensions and creates a Bell pair link within each macronode. The shown links are given by the adapted Bell pairs of Eq.\,\eqref{eq:GKPBellPairH}.}
	\label{fig:octo_3D_macronode}
\end{figure}

\begin{figure}
	\centering
	\includegraphics[]{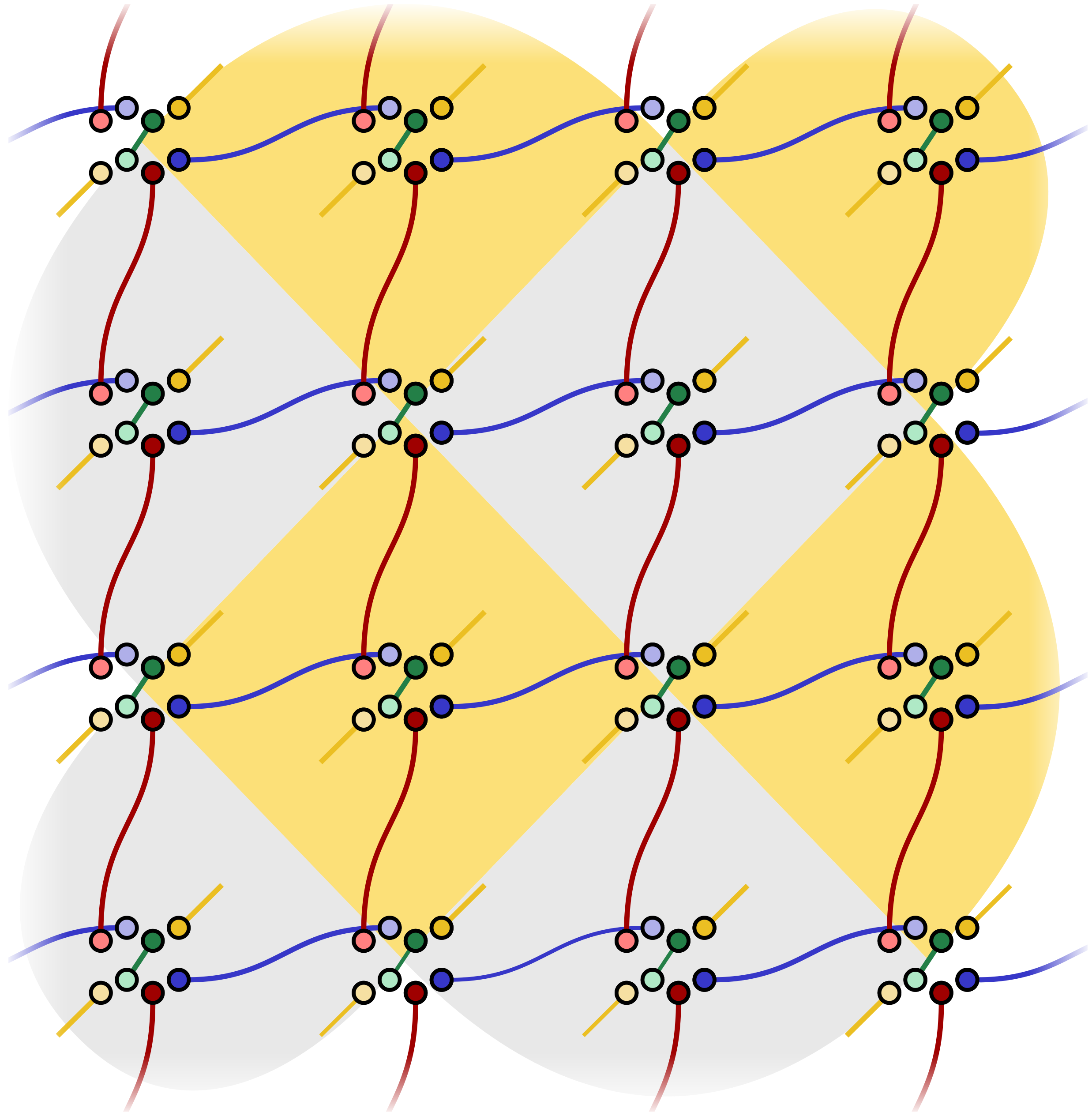}
	\caption{Correspondence of the first two dimensions of the ORL macronode lattice with the data and ancilla qubits of the planar, non-rotated surface code. The data qubits of the surface code sit on the vertices. The ancilla qubits in the center of the grey and yellow patches are used to measure the Z- and X-stabilisers of the code, respectively.}
	\label{fig:surface_code}
\end{figure}

While Clifford gates nicely illustrate the basic properties of the ORL, they can neither provide fault-tolerance nor universality, both of which can be achieved by running the surface code. This requires the continuous measurement of its stabilisers. To achieve this efficiently using the ORL, two modifications to the setup are required:
First, only three of the four dimensions are necessary to operate the planar surface code. Removing the longest delay line by setting $k=0$ results in a three-dimensional macronode lattice, where the corresponding GKP Bell pair has both its modes within the same macronode, forming an internal link. This three-dimensional ORL is depicted in Fig.\,\ref{fig:octo_3D_macronode}. Second, the input GKP Bell pairs need to be equipped with an additional Hadamard gate on one of their modes. In the optical setup this is easily implemented by a $\tfrac{\pi}{2}$-rotation and will be depicted by thick coloured lines between a light and a dark coloured circle:
\begin{align}\label{eq:GKPBellPairH}
    \includegraphics{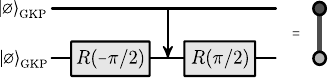}
\end{align}
This Hadamard gate acting on one half of a Bell pair is equivalent to a controlled-Z gate acting on two logical $\ket{+}$ states,
\begin{align}
    \bar H_1\left(\ket{00}+\ket{11}\right) = \bar H_2\left(\ket{00}+\ket{11}\right) = \bar C_Z\ket{++},
\end{align}
realising a standard DV cluster state \cite{PhysRevA.69.062311, PhysRevLett.86.910}. The resulting state created by the ORL is therefore a cubic lattice of macronodes connected by logical two-mode cluster states rather than Bell pairs. Among the three dimensions, one serves as the computational axis, while the remaining two form a square lattice of macronodes that function as data and ancilla qubits for the surface code. This structure is sketched in Fig.\,\ref{fig:surface_code}.

\begin{figure}
    \centering
    \includegraphics{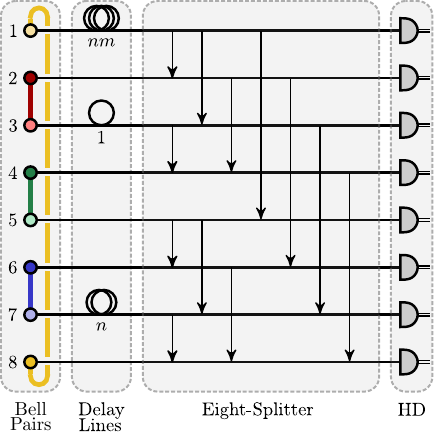}
    \label{fig:octo_3}
	\caption{Setup of the Octo-Rail Lattice configuration used to efficiently run the surface code. The GKP Bell pairs are equipped with an additional $\tfrac{\pi}{2}$-rotation, while the longest delay line is removed by setting $k=0$.}
\end{figure}

\subsection{Stabilisers}

\begin{figure}[h]
    \centering
    \begin{subfigure}[b]{\columnwidth}
    	\includegraphics[width=\columnwidth]{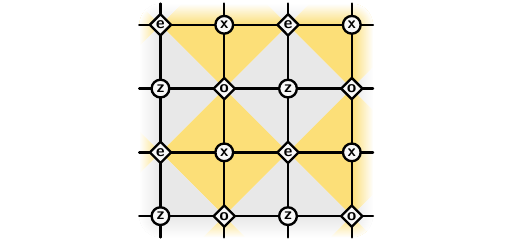}
    	\caption{Layout of the even and odd data qubits as well as ancilla qubits of the surface code. The symbols `e' and `o' represent the even and odd data qubits, respectively, while `z' and `x' denote the ancilla qubits related to the respective Z- and X-stabilisers of the surface code. Note that the measurement protocol is the same for all ancilla qubits, and the obtained surface code stabiliser depends only on the respective gates performed by neighboring data qubits.}
    	\label{fig:surfacelayout}
    \end{subfigure}
    \begin{subfigure}[b]{\columnwidth}
        \centering
        \includegraphics[width=0.5\columnwidth]{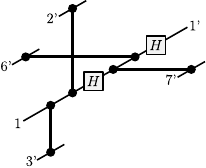}
        \caption{Circuit implementation of the even data qubit gates.}
        \label{fig:gate_even}
    \end{subfigure}
    \begin{subfigure}[b]{\columnwidth}
        \centering
        \includegraphics[width=0.5\columnwidth]{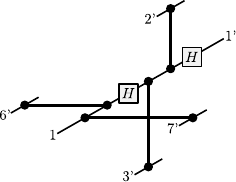}
        \caption{Circuit implementation of the odd data qubit gates.}
        \label{fig:gate_odd}
    \end{subfigure}
    \caption{Circuit diagrams for the gates implemented on macronodes corresponding to the even and odd data qubits of the surface code.}
    \label{fig:gate}
\end{figure}

The first step towards an implementation of the surface code is the measurement of its bulk stabilisers. Therefore, the ancilla qubits need to connect to their neighboring data qubits via controlled-Z and controlled-X gates, respectively, and subsequently be measured in the X basis. As the controlled-Z and -X gates are applied within the macronodes corresponding to the data qubits, it is convenient to describe the layout from the perspective of the data qubits. Each data qubit must be connected to the ancilla qubits by controlled-Z gates along one of the two dimensions and by controlled-X gates along the other.
This creates two types of data qubits: those connected to the Z-stabilisers of the surface code along the first dimension and its X-stabilisers along the second, referred to as “even”, and those connected to the X-stabilisers of the surface code along the first dimension and its Z-stabilisers along the second, referred to as “odd”. The gates that need to be performed within the even and odd data qubit macronodes are shown in Fig.\,\ref{fig:gate}. In contrast, all ancilla macronodes are identical and are measured in the X basis. Whether they correspond to a Z- or X-stabiliser of the surface code solely depends on their surrounding gate interactions. The layout of even and odd data as well as ancilla qubits is depicted in Fig.\,\ref{fig:surfacelayout}. The measurement bases required for one round of surface code stabilisers are,
\begin{align}\begin{split}\label{eq:angles_data}
    \text{even data: } &\left(0,0,0,\tfrac{\pi}{2},0,0,0,\tfrac{\pi}{2}\right)\\
    \text{odd data: } &\left(\tfrac{\pi}{2},0,0,0,0,0,0,\tfrac{\pi}{2}\right)\\
    \text{ancilla: } &\left(0,0,0,0,\tfrac{\pi}{2},0,0,\tfrac{\pi}{2}\right).
\end{split}\end{align}
The effect of these three gates on the quadratures of a macronode including noise propagation can be found in the supplementary material, but they can also be understood intuitively. In case of the even data qubits, measuring the same bases for the first and last four modes of the ORL removes one layer of beamsplitters effectively leaving two connected QRL macronodes. Measuring the first three modes in the Z basis ($\hat{x}$ quadrature, $\theta=0$) and the fourth in X ($\hat{p}$ quadrature, $\theta=\frac{\pi}{2}$) is then known to perform the two wanted controlled-Z gates followed by a teleportation through the GKP Bell pair with static Hadamard onto the connected macronode \cite{PRXQuantum.2.040353}. Specifically, following the notation of Fig.\,\ref{fig:octo_3D_macronode}, two controlled-Z gates are enacted connecting mode 1 with modes 2' and 3' of the neighboring macronodes, after which mode 1 is teleported to mode 5 including a GKP error correction as well as the Hadamard rotation. Combined with the same operation acting on the second half of the ORL macronode, the total gate ends up being the one given in Fig.\,\ref{fig:gate_even}. In case of the odd data qubits, the gate presented in Fig.\,\ref{fig:gate_odd} can simply be regarded as the gate for even data qubits preceded by the two transpositions $(26)$ and $(37)$ swapping modes 2 and 6, and 3 and 7, respectively. As this permutation is part of the group $P_\text{allowed}$, it only results in a permutation of measurement bases.

In order to understand the measurement of the ancilla qubits, one needs the relation of the quadratures going into the eightsplitter, $\vec{p}_\text{in} = (p_\text{in,1}, p_\text{in,2}, ... ,p_\text{in,8})$, and the ones coming out being measured in the homodyne detectors, $\vec{p}_\text{m}= (p_\text{m,1}, p_\text{m,2}, ... ,p_\text{m,8})$. It is given by the matrix
\begin{align}
    S = \frac{1}{2\sqrt{2}}\begin{pmatrix}
        1& -1& -1&  1& -1&  1&  1& -1 \\
        1&  1& -1& -1& -1& -1&  1&  1 \\
        1& -1&  1& -1& -1&  1& -1&  1 \\
        1&  1&  1&  1& -1& -1& -1& -1 \\
        1& -1& -1&  1&  1& -1& -1&  1 \\
        1&  1& -1& -1&  1&  1& -1& -1 \\
        1& -1&  1& -1&  1& -1&  1& -1 \\
        1&  1&  1&  1&  1&  1&  1&  1 
    \end{pmatrix}
\end{align}
with $\vec{p}_\text{m}=S\vec{p}_\text{in}$. Note that  $S$  is only a quarter of a symplectic matrix as  $x$- and  $p$-quadratures remain decoupled, and the focus is solely on measurements in the  X  basis. Consequently, measuring the $p$-quadrature of modes 5 and 8 and the $x$-quadrature of all others gives exactly the desired product of X-stabilisers with
\begin{align}
    p_{\text{m}, 8} - p_{\text{m}, 5} \propto  p_{\text{in}, 2} + p_{\text{in}, 3} + p_{\text{in}, 6} + p_{\text{in}, 7}.
\end{align}
Besides, it is also possible to measure two-mode stabilisers required for the boundaries of the (anticlockwise) rotated surface code \cite{PhysRevA.76.012305}. Here, the measurement bases
\begin{align}\label{eq:angles_boundary1}
    \left(0,0,0,0,\tfrac{\pi}{2},\tfrac{\pi}{2},\tfrac{\pi}{2},\tfrac{\pi}{2}\right)
\end{align}
separate the stabilisers top left and bottom right, creating a vertical edge, as
\begin{align}\begin{split}
    (p_{\text{m}, 8} - p_{\text{m}, 5}) + (p_{\text{m}, 6} - p_{\text{m}, 7}) &\propto  p_{\text{in}, 2} + p_{\text{in}, 6}\\
    (p_{\text{m}, 8} - p_{\text{m}, 5}) - (p_{\text{m}, 6} - p_{\text{m}, 7}) &\propto  p_{\text{in}, 3} + p_{\text{in}, 7}
\end{split}\end{align}
while the measurement bases
\begin{align}\label{eq:angles_boundary2}
    \left(0,\tfrac{\pi}{2},\tfrac{\pi}{2},0,\tfrac{\pi}{2},0,0,\tfrac{\pi}{2}\right)
\end{align}
separate the stabilisers bottom left and top right, creating a horizontal edge, given that
\begin{align}\begin{split}
    (p_{\text{m}, 8} - p_{\text{m}, 5}) + (p_{\text{m}, 2} - p_{\text{m}, 3}) &\propto  p_{\text{in}, 2} + p_{\text{in}, 7}\\
    (p_{\text{m}, 8} - p_{\text{m}, 5}) - (p_{\text{m}, 2} - p_{\text{m}, 3}) &\propto  p_{\text{in}, 3} + p_{\text{in}, 6}.
\end{split}\end{align}
In general, the presented architecture is best suited to run the more efficient rotated version of the surface code as the skewed periodic boundaries allow an efficient use of resources despite rotating patches by $45^\circ$. Together, the macronode measurements listed in Eqs.\,(\ref{eq:angles_data},\,\ref{eq:angles_boundary1},\,\ref{eq:angles_boundary2}) are sufficient to maintain a given patch of the rotated surface code on the ORL.

\subsection{Universality}

\begin{figure}[h!]
    \centering
    \begin{subfigure}[b]{\columnwidth}
    	\includegraphics[width=\columnwidth]{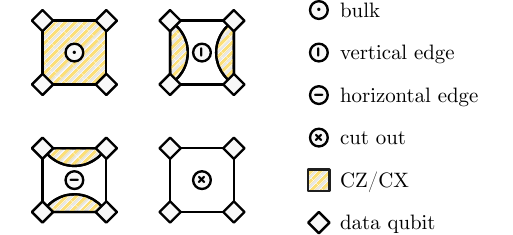}
    	\caption{Types of ancilla qubits.}
    	\label{fig:patches_ancilla}
    \end{subfigure}
    \begin{subfigure}[b]{\columnwidth}
        \centering
        \includegraphics[width=\columnwidth]{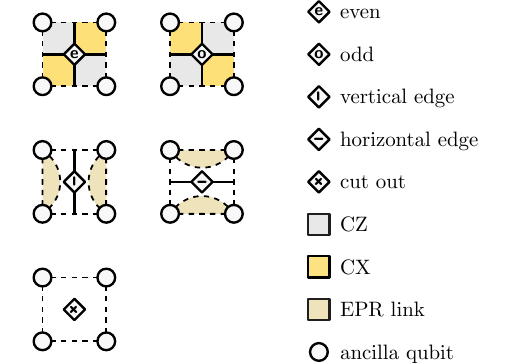}
        \caption{Types of data qubits.}
        \label{fig:patches_data}
    \end{subfigure}
    \begin{subfigure}[b]{\columnwidth}
        \centering
        \includegraphics[width=\columnwidth]{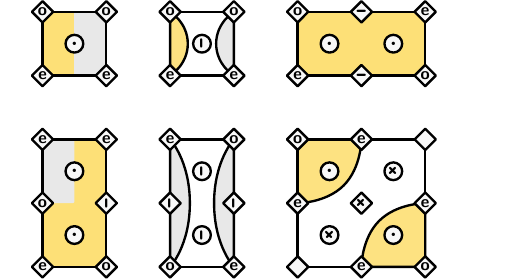}
        \caption{Examples of stabilizers arising from their composition.}
        \label{fig:patches_stabilizers}
    \end{subfigure}
    \caption{Building blocks required for universality \cite{Litinski_2019}. The measurement bases necessary to run an arbitrary computation with the surface code are split into types of ancilla qubits (a) and data qubits (b). Their correct combination allows the implementation of a variety of stabilizers (c). The shown examples are a mixed bulk, mixed edge and double stabilizer, as well as a twist defect, double edge and corner stabilizer.}
    \label{fig:SurfaceCodePatches}
\end{figure}

Storing logical qubits in patches of the rotated surface code not only provides fault-tolerance, but can also be used to perform any universal quantum computation. In \cite{Litinski_2019} this is achieved by breaking down a given algorithm into magic Pauli product rotations of the form,
\begin{align}\label{eq:magicPauliRotations}
    \exp\left(-i\frac{\pi}{8}\hat P_1\otimes \hat P_2\otimes...\otimes \hat P_n\right),
\end{align}
with $\hat P_i\in\{\mathbbm{1}, \hat X, \hat Y, \hat Z\}$, together with Clifford gates that can be performed in post-processing. Their implementation requires the initialisation of surface code patches in certain logical states as well as the ability to measure a selection of stabilizers that go beyond the simple bulk and edge stabilizers.

\paragraph{Patch initialisation}
In order to achieve universality, one needs to be able to initialise surface code patches in the basis states $\ket{0}$ and $\ket{+}$ as well as the magic state $\ket{T}\propto \ket{0} + e^{i\pi/4}\ket{1}$. Initialisation of the two basis states can be achieved by preparing all data qubits in the desired state within the GKP encoding and subsequently measuring the bulk and edge stabilizers of the patch. Measuring all modes of a data qubit in X-basis, corresponding to measurement bases
\begin{align}\label{eq:angles_init0}
    \left(\tfrac{\pi}{2},\tfrac{\pi}{2},\tfrac{\pi}{2},\tfrac{\pi}{2},\tfrac{\pi}{2},\tfrac{\pi}{2},\tfrac{\pi}{2},\tfrac{\pi}{2}\right),
\end{align}
provides a $\ket{0}_\text{GKP}$ state within the next surface code layer, while measuring all modes in Z-basis, corresponding to measurement bases
\begin{align}\label{eq:angles_init+}
    \left(0,0,0,0,0,0,0,0\right),
\end{align}
results in a $\ket{+}_\text{GKP}$ state within the next layer.

\begin{figure}
    \centering
    \includegraphics{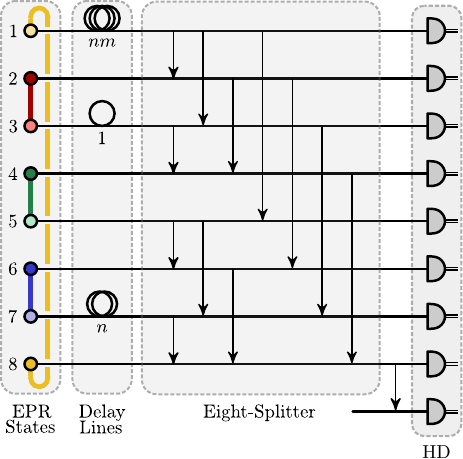}
    \caption{Adaptation of the ORL setup providing universality. By adding a balanced beamsplitter with vacuum input and one homodyne detector on mode eight, heterodyne detection can now be used to generate GKP magic states within the cluster.}
    \label{fig:ORL_heterodyne}
\end{figure}

Magic state initialisation, on the other hand, requires the more complex and resource-intensive process of magic state distillation \cite{Litinski_2019,Litinski2019magicstate}. To this end, single $\ket{T}_\text{GKP}$ states are grown into separate surface code patches via a non-fault-tolerant state injection procedure \cite{Horsman_2012, landahl2014quantumcomputingcolorcodelattice}. Afterwards, a multitude of these faulty patches can be distilled into the desired high-quality logical magic state. The initial $\ket{T}_\text{GKP}$ state can be obtained by using heterodyne detection on one half of a GKP Bell pair \cite{PhysRevLett.123.200502}. This can easily be incorporated into the existing setup by replacing the homodyne detector of mode eight with a heterodyne detector, i.e. a balanced beamsplitter mixing mode eight with vacuum followed by two homodyne detectors measuring conjugate quadratures, see Fig.~\ref{fig:ORL_heterodyne}. Measuring modes one to seven of a data macronode as well as the four surrounding ancilla qubits in Z-basis then isolates the GKP Bell pair of mode eight and allows for the initialisation of GKP magic states within the cluster. On the other hand, homodyne detection on mode eight is still possible by just measuring the same quadrature on both detectors.

In general, the outcome of a teleportation through a GKP Bell pair changes with the results of the two homodyne detectors. In case of the $\ket{0}_\text{GKP}$ and $\ket{+}_\text{GKP}$ state initialisation, this may lead to an additional Pauli operation, while in magic state initialisation, this may result in lower quality states. Possible improvements are non-orthogonal measurement angles of the heterodyne detector effectively substituting the vacuum with a more general Gaussian \cite{PRXQuantum.5.020337}, and replacing more than one homodyne detector by heterodyne detectors.

\paragraph{Stabilizer measurements}
Any given magic Pauli product rotation can be performed by measuring the desired product of Pauli operators multiplied with the Z operator of a magic ancilla qubit. Within the surface code, this can be achieved by initialising a magic $\ket{T}$ patch and connecting its Z-edge to the respective Z- and X-edges of different qubit patches using a non-square ancilla \cite{Litinski_2019}.
On one hand, this requires the measurement of bulk and edge stabilizers needed for general qubit patches. On the other hand, the connecting ancilla patch requires more complex stabilizer measurements also found in twist-based lattice surgery \cite{PhysRevX.7.021029, Litinski2018latticesurgery}.
Namely, double sized bulk and edge stabilizers are necessary to align the edge of the ancilla patch with both the Z- and X-edges of the qubit patches. These double stabilizers shown in Fig.~\ref{fig:patches_stabilizers} skip either one row or one column of data qubits, spanning a total of six data qubits, and can be used to offset the check pattern of Z- and X-stabilizers. 
They can be implemented by measuring the skipped data qubits in the measurement bases
\begin{align}\label{eq:angles_double1}
    \left(\tfrac{\pi}{2},\tfrac{\pi}{2},\tfrac{\pi}{2},\tfrac{\pi}{2},0,0,0,0\right),
\end{align}
to create a vertical Bell pair and skip one row or in
\begin{align}\label{eq:angles_double2}
    \left(\tfrac{\pi}{2},0,0,\tfrac{\pi}{2},0,\tfrac{\pi}{2},\tfrac{\pi}{2},0\right)
\end{align}
to create a horizontal Bell pair skipping one column. The double stabilizer is then easily obtained by multiplying the respective measurement outcomes of its two ancilla qubits.

Similarly, a twist defect stabilizer, which mediates between one double sized and two normal sized stabilizers, can be implemented by skipping only one of the two middle data qubits, see Fig.~\ref{fig:patches_stabilizers}. Multiplying the measurement outcomes of its two ancilla qubits then provides a weight-five stabilizer that includes the $Y\propto Z\cdot X$ information of the remaining middle data qubit and allows to measure the combined Z- and X-edge of a patch.
Finally, both data and ancilla qubits can be cut out of the cluster state, removing all connections to other macronodes, by measuring all its modes in Z,
\begin{align}\label{eq:angles_cut}
    \left(0,0,0,0,0,0,0,0\right).
\end{align}
Together, the measurement bases of Eqs.\,(\ref{eq:angles_data},\,\ref{eq:angles_boundary1},\,\ref{eq:angles_boundary2},\,\ref{eq:angles_init0},\,\ref{eq:angles_init+},\,\ref{eq:angles_double1},\,\ref{eq:angles_double2}) can then measure all stabilisers needed to perform any multi-qubit magic Pauli product rotation on the surface code as described in \cite{Litinski_2019} and render the presented ORL architecture universal. A list of required ancilla and data qubit measurements, as well as some stabilizers resulting from their combination can be found in Fig.~\ref{fig:SurfaceCodePatches}.

Notably, the architecture of Fig.~\ref{fig:ORL_heterodyne} does not inherently require pre-generated GKP magic states and all necessary real-time feed-forward operations can be implemented through adjustments to the measurement bases of the nine homodyne detectors. The primary challenge, therefore, lies in the generation of the eight GKP qunaught states required as inputs. Although some progress has been made in their generation at optical frequencies \cite{konno2024logical, Larsen2025}, this step remains a substantial technical hurdle in the implementation of the system.

\subsection{Fault-tolerance Threshold}

The main purpose of the surface code is the provision of fault-tolerance to the architecture. This comes at the cost of a large overhead of physical qubits needed per logical qubit. How beneficial the surface code encoding is depends on the error rate of these individual physical qubits. For very noisy physical qubits, increasing the size of a surface code patch will result in a total increase of noise and worsen the logical qubit. On the other hand, for low physical error rates, increasing the patch size will decrease the logical error rates resulting in logical qubits with arbitrarily low noise. The break-even point, where increasing the size of a surface code patch leaves its logical error rate unchanged, is known as the fault-tolerance threshold. For GKP qubits acting as physical qubits, their physical error rate is given by their squeezing and the break-even point is consequently referred to as the squeezing threshold. As the difficulty of the experimental realisation of qunaught GKP states needed as inputs only increases with higher squeezing, this squeezing threshold acts as a good benchmark for how efficiently a system uses its resources.

\begin{figure}[h]
    \centering
    \includegraphics[width=\columnwidth]{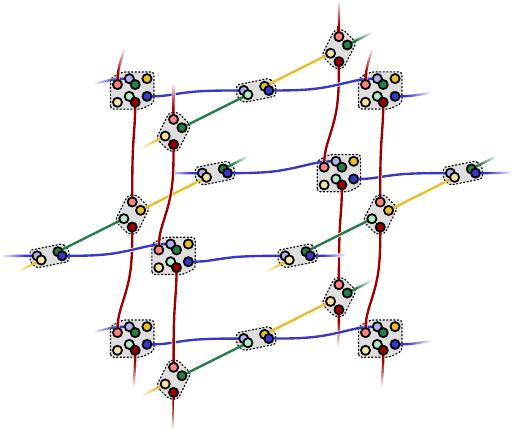}
    \caption{Implementation of the RHG lattice on the ORL. The measurement bases used for even and odd data qubits effectively split them into two 4-mode macronodes connected by their internal link. Performing this split graphically quickly reveals the underlying RHG lattice. Shown are one and a half layers of surface code.}
    \label{fig:RHG}
\end{figure}

Simulations of a Raussendorf-Harrington-Goyal (RHG) \cite{Raussendorf_2007} lattice of macronodes running the surface code have established a squeezing threshold of 9.75 dB using a two-stage minimum-weight perfect-matching decoder fed by results from the underlying GKP measurements \cite{PRXQuantum.2.040353, aghaee2025scaling, noh2022low}.
This RHG lattice is a three-dimensional topological structure central to fault-tolerant quantum computation. In MBQC, the RHG lattice embodies the planar surface code, enabling logical operations through topological manipulations of encoded qubits. Within the ORL cluster state, the measurement angles presented above effectively split each macronode into two parts which are connected by the internal link. Treating these two parts of each macronode separately transforms the ORL into a RHG lattice consisting of QRLs, as shown in Fig.~\ref{fig:RHG}. This is exactly the setup simulated in \cite{PRXQuantum.2.040353, aghaee2025scaling} and the squeezing threshold of 9.75 dB also applies to the ORL. While the two designs are structurally equivalent, the one presented in \cite{PRXQuantum.2.040353, aghaee2025scaling} uses two spatial and one temporal dimension compared to the fully temporal encoding of the ORL. By replicating the connectivity and noise propagation required to run the surface code efficiently, the ORL establishes itself as a viable and near-term scalable platform for fault-tolerant quantum computation.

\section{Adapting and Extending the ORL}

\begin{figure}
	\centering
	\includegraphics[width=0.68\columnwidth]{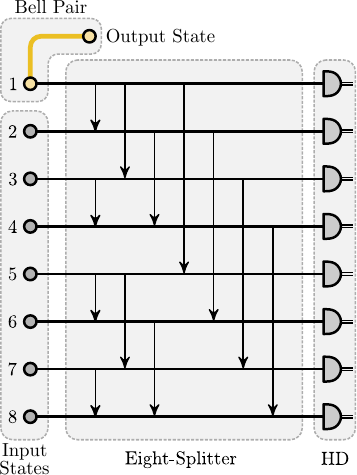}
	\caption{Adaptation of the ORL useful for multiplexing. Each of the seven inputs can be teleported to the output state by changing the measurement bases of the homodyne detectors.}
	\label{fig:Multiplexing}
\end{figure}
\begin{figure}
	\centering
	\includegraphics[width=0.8\columnwidth]{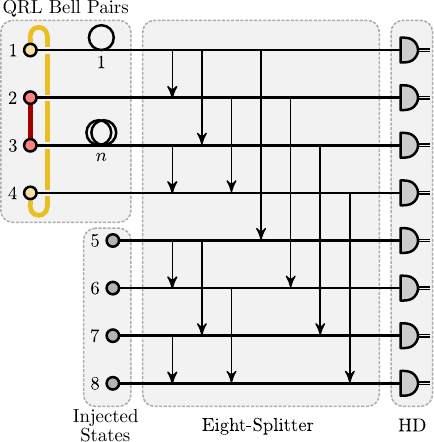}
	\caption{Adaptation of the ORL useful for state injection. Half of the macronode is used to operate the cluster state, while the other half provides input states that can be injected when needed.}
	\label{fig:StateInjection}
\end{figure}

A unique selling point of the ORL is its small setup size and ease of near-term scalability combined with enough flexibility to enable fault-tolerant and universal quantum computation. This originates from its temporal, rather than spatial encoding, allowing the progression from the two-dimensional QRL to the four-dimensional ORL with only a linear increase in setup size. Moreover, this procedure can be generalised to $2^n$ dimensions, where $n\in\mathbb{N}$. Notably, the setup size scales linearly with the number of dimensions requiring $2^{n+1}$ input states and homodyne detectors, while the number of beamsplitters a mode passes before measurement only scales logarithmically requiring $n+1$ layers of $2^n$ beamsplitters. A cluster state of arbitrary dimension can be achieved by removing some of the delay lines as seen in the surface code implementation. While this might seem inefficient, the use of the symmetric beamsplitter network guarantees that all operations of the DRL, the QRL, as well as the ORL can still be performed with the same amount of noise. Note that this is based on the assumption of ideal beamsplitters and homodyne detectors. However, given the high efficiency of both beamsplitters and homodyne detectors in optical setups together with the low number of passes for each individual mode even for high dimensions, this assumption seems reasonable.

Besides the extension to higher dimensions, there are several adaptations to the ORL that can be considered, most of which have already been introduced: Different arrangements of inputs, corresponding to cosets of the group of permutations, can be used to run specific gates more efficiently. Replacing homodyne with heterodyne detectors opens the door for magic state generation. The adaptation of the GKP Bell pairs, however, has only been briefly discussed. While adding rotations to the input states already facilitated the implementation of certain gates, modifying the Bell pairs themselves can give rise to different underlying GKP encodings. Any general GKP encoding is related to the square encoding by a Gaussian transformation with codewords given by
\begin{align}
    \ket{j}_\text{GKP$^\prime$}=\hat U_{G}\ket{j}_\text{GKP}.
\end{align}
As identical single-mode Gaussian operations excluding displacements commute with a beamsplitter, the Bell pair in the desired encoding can be generated by modifying the GKP qunaught states
\begin{align}
    \ket{00}_\text{GKP$^\prime$}+\ket{11}_\text{GKP$^\prime$}\propto \hat B_{12}\left(\hat U_{G}\ket{\varnothing}\right)_1\left(\hat U_{G}\ket{\varnothing}\right)_2.
\end{align}
When teleporting through this new Bell pair, the enacted operator becomes
\begin{align}
    \hat N(\beta)\hat U_{G}\hat \Pi_\text{GKP}\hat U^T_{G}\hat N(\beta)=\hat N(\beta)\hat \Pi_{\text{GKP}'}\hat U_{G}\hat U^T_{G}\hat N(\beta),
\end{align}
where the transpose is defined by its effect on the quadratures $\hat x^T=\hat x$ and $\hat p^T=-\hat p$ \cite{PhysRevA.102.062411}.
Besides, the applied gates now also act within the new GKP code space. In order to apply the same logical gates as in the square encoding, the application of
\begin{align}
    \hat V_{\text{GKP}'}\left(\theta_1, \theta_2\right)=\hat U_G \hat V\left(\theta_1, \theta_2\right)\hat U_G^\dagger
\end{align}
is required. Fortunately, for every $\theta_1$, $\theta_2$ there exist angles $\phi_1$, $\phi_2$ so that
\begin{align}\label{eq:Vtranspose}
    \hat V\left(\phi_1,\phi_2\right)=\hat U^{\dagger T}_{G}\hat V\left(\theta_1,\theta_2\right)\hat U^\dagger_{G}
\end{align}
and we can write
\begin{align}
    \hat V\left(\phi_1, \phi_2\right)=\left(\hat U_G\hat U^T_G\right)^\dagger \hat V_{\text{GKP}'}\left(\theta_1, \theta_2\right).
\end{align}
The derivation of the two angles $\phi_1(\theta_1, \theta_2)$ and $\phi_2(\theta_1, \theta_2)$ can be found in the supplementary material.
As a result, any logical gate which can be performed in the square encoding can also be implemented in the arbitrary GKP encoding when accepting the modified noise term
\begin{align}
    \hat U_G\hat U^T_G\hat N(\beta)\left(\hat U_G\hat U^T_G\right)^\dagger.
\end{align}
For rectangular and hexagonal encodings, this noise modification can be reduced to a logical operator as $\hat U_\text{rec}\hat U^T_\text{rec}=\bar I_\text{rec}$ and $\hat U_\text{hex}\hat U^T_\text{hex}=\bar H_\text{hex}$, respectively. Consequently, both encodings can be run on the ORL architecture without changing the underlying noise distribution.
Static gates enacted directly on the Bell pairs also need to be changed. In case of the Hadamard gate, this turns a simple rotation into an active operation that involves squeezing. If the macronode lattice is bipartite, this can be accounted for dynamically through a change of measurement bases \cite{walshe2024}, otherwise the gate has to be implemented statically. That is the case for the presented surface code implementation with $k=0$, as only odd $n$, $m$ and $k$ would provide a bipartite macronode lattice.

\begin{figure}
	\centering
	\includegraphics[width=0.55\columnwidth]{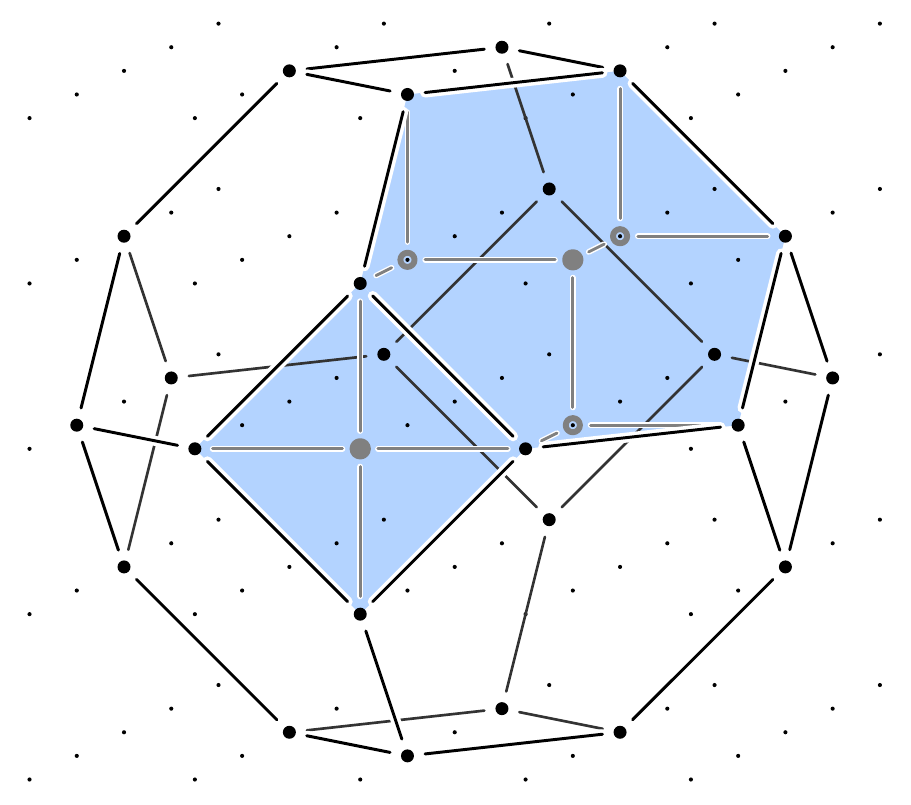}
	\caption{Bulk stabilisers of the three-dimensional gauge color code on a cubic lattice including ancilla qubits needed for the stabiliser measurements.}
	\label{fig:ColorCode}
\end{figure}
In general, the progression from DRL to QRL to ORL, and beyond, enables the implementation of high-dimensional topological error correction codes combined with arbitrary GKP encodings, providing an excellent platform for exploring advanced error correction schemes beyond the standard surface code. For example, the gauge color code \cite{Bombín_2015} seen in Fig.\,\ref{fig:ColorCode} can be implemented on the ORL, yet is difficult to implement in spatial setups due to its three-dimensional layout. Similarly, the four-dimensional surface code \cite{alicki2008, 10.1063/1.1499754}, which is known to showcase some advantageous properties over its two-dimensional counterpart, can be realised on a higher-dimensional extension of the ORL but cannot be implemented in a spatial setup without non-local interactions or requiring costly rearrangements of qubits. Rectangular GKP codes can be used to bias the noise, enabling the implementation of tailored adaptations of the surface code such as the XZZX surface code \cite{xzzx, PhysRevA.108.052428}. Additionally, concatenation of different codes, which requires higher-dimensional connectivity, provides a promising avenue for enhancing fault-tolerance thresholds.

Besides performing fault-tolerant quantum computation, there is another natural use case for the ORL extensions. Due to an easily scalable number of input modes and low-noise identity gates that can be performed between all of them, they are well suited for multiplexing. More specifically, replacing all Bell pairs but one with input states, the measurement bases of the homodyne detectors determine which input state is teleported to the output state connected by the remaining Bell pair. In this way, the accumulated noise is the same as for the DRL and thus independent of the number of input modes. When a GKP error correction is not wanted, the GKP Bell pair can be replaced by a two-mode squeezed state. This setup is depicted in Fig.\,\ref{fig:Multiplexing}. Furthermore, this multiplexing setup can be combined with the cluster state generation enabling the state injection into a fully functional cluster state. This is shown in Fig.\,\ref{fig:StateInjection} for the combination of a QRL and a four-mode multiplexer.
\section{Conclusions}\label{sec:Conclusions}

The Octo-Rail Lattice (ORL) introduced in this work combines the benefits of measurement-based quantum computation with topological quantum error correction in an experimentally practical photonic architecture. By extending the entanglement structure of previously developed Dual-Rail and Quad-Rail Lattices into four dimensions, the ORL achieves fault-tolerance and universality within a compact and static physical footprint composed entirely of passive optical elements such as beamsplitters, delay lines, and homodyne detectors. Crucially, no active optical switching is required, and the ORL simultaneously reduces the spatial overhead commonly associated with previous proposals, which greatly reduces implementation complexity and positions the ORL as a feasible near-term candidate for scalable quantum computing, free from the implementation challenges inherent to spatial scaling.

Leveraging Gottesman-Kitaev-Preskill (GKP) qunaught states, the ORL performs Clifford gate teleportation and GKP error correction within each measurement step, where adaptive homodyne measurements implement quantum logic operations while continuously correcting quadrature displacement errors intrinsic to CV systems. Additionally, the compatibility of the ORL architecture with arbitrary GKP encodings further broadens its potential applications, accommodating use cases involving biased noise or hardware specific encodings. The three-dimensional structure of the reduced ORL naturally implements the surface code, specifically reproducing the established Raussendorf-Harrington-Goyal (RHG) lattice, thereby providing topological protection against logical errors. The presented single-step approach simplifies the computational protocol, reduces the accumulation of gate noise, and minimises requirements for fault-tolerance thresholds, making the ORL architecture compatible with the existing fault-tolerance threshold of 9.75 dB squeezing.

By employing heterodyne-based measurements for magic state preparation, universality can be ensured without additional non-Gaussian resources or complex feedforward control beyond the adaptive homodyne measurements. Thus, the experimental challenge is concentrated in the reliable generation of the highly non-classical GKP input states.

Beyond its implementation of the surface code, the ORL generalises naturally to cluster states of higher dimension, extending its capabilities to more advanced topological error correction schemes, such as three-dimensional gauge color codes or four-dimensional topological codes, without substantial modification to the physical setup. Crucially, the symmetry and balanced splitting ratios of the ORL beamsplitter network ensure that scaling up to higher dimensions does not introduce significant additional optical noise. Furthermore, because each additional dimension is encoded using time-domain multiplexing via fixed length fibre delays, the hardware complexity grows only linearly with the number of dimensions. Additionally, this structure facilitates the efficient integration of mode multiplexing, further enhancing the versatility of the ORL architecture for broader quantum computing and communication applications.

Besides, the ORL design offers substantial advantages for state injection, critical for implementing universal quantum logic and supporting quantum communication networks. The passive structure of the ORL, combined with the flexibility of adaptive local homodyne measurement bases, enables selective routing and efficient switch-free injection of input states such as magic states or encoded quantum information. 

Ultimately, the ORL provides a realistic and experimentally viable pathway toward large-scale photonic quantum computing. By combining simplicity in optical design, flexibility in logical operations, and inherent static nature, it directly addresses long-standing practical challenges in the field. The ORL thus bridges the gap between theoretical CV quantum computing frameworks and tangible experimental realisation, setting a clear trajectory for further advancements. It is anticipated that the scalability of this framework will inspire new experimental approaches, allowing for the exploration of a large variety of error correction codes, and further advancements in fault-tolerant quantum computing architectures. 
\begin{acknowledgments}
We gratefully acknowledge support from
the Danish National Research Foundation, Center for Macroscopic Quantum States (bigQ,\ DNRF0142), 
EU project CLUSTEC (grant agreement no. 101080173),
EU ERC project ClusterQ (grant agreement no. 101055224, ERC-2021-ADG), NordicQuEst (grant agreement no. CF21-0657)
Innovation Fund Denmark (PhotoQ project, grant no.\ 1063-00046A),
and funding from the BMBF in Germany (QR.X/QR.N, QuKuK, QuaPhySI, PhotonQ).
\end{acknowledgments}
\appendix
\section{Background}\label{sec:Background}

In this section, the general concepts behind CV macronode cluster states as well as the GKP code, which form the basis of the paper, are introduced. Additionally, a brief review of topological quantum error correction codes is provided, with a specific focus on the two-dimensional surface code, which is used to demonstrate the universality and fault-tolerance of the presented architecture. 

\subsection{Notation and Conventions}\label{sec:Notation}

Throughout the paper, the convention $\hbar = 1$ are used. The canonical quadrature operators of a CV mode are thus given by $\hat x = \frac{1}{\sqrt{2}}\left(\hat a + \hat a ^\dagger\right)$ and $\hat p = \frac{1}{\sqrt{2}i}\left(\hat a - \hat a ^\dagger\right)$ resulting in the commutator $\left[\hat x, \hat p\right]=i$ and a vacuum variance of $\frac{1}{2}$. The eigenstates of these operators, commonly referred to as infinitely squeezed states, will be denoted by, \begin{align}
    \hat x \ket{a}_x = a\ket{a}_x &&\text{and}&& \hat p \ket{b}_p = b\ket{b}_p,
\end{align}
where $a, b\in\mathbb{R}$ are the eigenvalues associated with $\hat{x}$ and $\hat{p}$, respectively.
The set of Gaussian operations is generated by the phase-rotation operator,
\begin{align}
    \hat R(\theta)=\exp\left(i\theta\hat a^\dagger\hat a\right),
\end{align}
the displacement operator,
\begin{align}
    \hat D(x_0+ip_0)=\exp\left(i\sqrt{2}\left(p_0\hat x-x_0\hat p\right)\right),
\end{align}
the squeezing operator,
\begin{align}
    \hat S(t)=\exp\left(\frac{i}{2}\ln(t)\left(\hat x\hat p+\hat p\hat x\right)\right),
\end{align}
for $t>0$ (where ln(t) is the standard squeezing parameter), as well as the beam-splitting operation between modes $j$ and $k$,
\begin{align}
    \hat B_{jk}(\varphi)=\exp\left(i\varphi\left(\hat p_j\hat x_k-\hat x_j\hat p_k\right)\right).
\end{align}
Special cases of these operators, which will be used throughout this paper, are the Fourier operator $\hat F=\hat R\left(\frac{\pi}{2}\right)$ and the balanced beamsplitter $\hat B_{jk}=\hat B_{jk}\left(\frac{\pi}{4}\right)$. 
Homodyne measurements in the rotated quadrature,
\begin{align}
    \hat x_\theta=\hat R(\theta)\hat x\hat R^\dagger(\theta)=\hat x\cos\theta +\hat p \sin\theta, 
\end{align}
will be denoted by,
\begin{align}
    \prescript{}{x_\theta}{\bra{m}}=\prescript{}{x}{\bra{m}}\hat R^\dagger(\theta),
\end{align}
with the measurement outcome $m$.

\subsection{The Gottesman-Kitaev-Preskill Code}\label{sec:GKP}

Bosonic codes are necessary to achieve fault-tolerant computation with continuous variables as they enable the encoding of discrete quantum information within the infinite dimensional Hilbert spaces of bosonic modes. The most promising candidate is the GKP code \cite{GKP} due to its excellent performance under photon loss \cite{PhysRevA.97.032346, zheng2024performance} and phase noise \cite{Leviant2022quantumcapacity} -- the two main sources of error in optical setups -- as well as its straightforward compatibility with Gaussian cluster states and homodyne detection \cite{PhysRevA.102.062411}. Its ideal basis states are given by,
\begin{align}
    \ket{j}_\text{GKP}=\sum_{s\in\mathbb{Z}}\ket{\sqrt{\pi}(2s+j)}_x =\sum_{s\in\mathbb{Z}}(-1)^{js}\ket{\sqrt{\pi}s}_p,
\end{align}
with $j=0, 1$. While this is known as the square encoding, general GKP codes can be obtained by applying a Gaussian transformation to the basis states. Noteworthy alternatives are the rectangular codes with,
\begin{align}
    \ket{j_\alpha}_\text{rec}=\hat S\left(\frac{\sqrt{\pi}}{\alpha}\right)\ket{j}_\text{GKP},
\end{align}
which can be favourable in case of biased noise \cite{PhysRevA.108.052428}, as well as the hexagonal encoding given by,
\begin{align}
    \ket{j}_\text{hex}=\hat S\left(\sqrt[4]{3}\right)\hat R\left(\frac{\pi}{4}\right)\ket{j}_\text{GKP}.
\end{align}
A Clifford gate set for the square GKP code consisting of Hadamard $(\bar H)$, phase $(\bar P)$ and controlled-Z gate $(\bar C_Z)$ can be generated by the Gaussian operations
\begin{align}
    \bar H=\hat F,\qquad \bar P=\hat P(-1),\qquad \bar C_Z=\hat C_Z(1),
\end{align}
with the shearing operator,
\begin{align}
    \hat P(\sigma)=\exp\left(\frac{i}{2}\sigma\hat x^2\right)\stackrel{\sigma>0}{=}\hat R\left(-\gamma\right)\hat S\left(\cot\gamma\right)\hat R \left(-\gamma-\frac{\pi}{2}\right),
\end{align}
where $\gamma=\frac{1}{2}\atan\left(\frac{2}{\sigma}\right)$, and the controlled-Z gate,
\begin{align}\label{eq:CZGateDecomposition}
    \hat C_{Z,jk}(g)=\exp\left(ig\hat x_j\hat x_k\right)=\hat B_{kj}\hat P_j(-g)\hat P_k(g)\hat B_{jk}.
\end{align}
The logical information of a given GKP qubit can be accessed in the X, Y and Z Pauli bases by homodyne measurements of the $\hat x$, $\hat x_\frac{\pi}{4}$ and $\hat p$ quadratures, respectively. In the presence of noise, the measurement outcomes may no longer lie on the ideal GKP grid and need to be assigned to the closer of the two basis states. This becomes especially relevant when considering realistic GKP states.
As ideal code states are unphysical, due to their infinite energy, it is necessary to consider finite energy approximations. An especially symmetric and easily workable set of physical GKP states can be obtained by using the non-unitary damping operator,
\begin{align}
    \hat N(\beta)=\exp\left(-\beta\hat n\right).
\end{align}
The resulting states, commonly referred to as approximate GKP states, are given by,
\begin{align}
\begin{split}
    \hat N(\beta)\ket{j}_\text{GKP}\propto &\int\hspace{-0.7mm} dx\sum_{s\in\mathbb{Z}}\exp\left(-\frac{\Delta^2}{2}\left(\sqrt{\pi}(2s+j)\right)^2\right.\\&\left.-\frac{1}{2\Delta^2}\left(x-\sqrt{1-\Delta^4}\sqrt{\pi}(2s+j)\right)^2\right)\ket{x}.
\end{split}
\end{align}
with $\Delta^2=\sinh(\beta)$ \cite{PhysRevA.102.032408}. In contrast to their ideal counterpart, the approximate GKP states exhibit an infinite sum of Gaussian peaks weighted by an overall Gaussian envelope. The variance of the Gaussian peaks as well as the Gaussian envelope are determined by the parameter $\Delta^2$, which is known as the squeezing of the state and commonly quoted in decibels,
\begin{align}
    \left(\Delta^2\right)_\text{dB}=-10\cdot\log_{10}\left(\Delta^2\right).
\end{align}
Due to the infinite support of their Gaussian peaks, the approximate GKP basis states are not orthogonal, i.e.
\begin{align}
    \prescript{}{\text{GKP}}{\bra{0}}\hat N^\dagger(\beta)\hat N(\beta)\ket{1}_\text{GKP}\gneq0.
\end{align}
When accessing their logical information, this may lead to a misrepresentation of the homodyne outcome and a subsequent logical error. The probability of these logical errors decreases exponentially with the level of squeezing \cite{GKP},
\begin{align}
    P_\text{error}\simeq\frac{2\Delta}{\pi}e^{-\frac{\pi}{4\Delta^2}}.
\end{align}

\subsection{Generalised Teleportation and Knill Error Correction} \label{sec:Teleportation}

The straightforward compatibility of the GKP code with CV cluster states is based on the close relation of CV quantum teleportation with the Knill error correction \cite{PhysRevA.102.062411} for approximate GKP states. The former is described by the circuit
\begin{align}
    \includegraphics{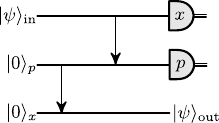}.
\end{align}
First, two squeezed states are sent through a beamsplitter, creating a two-mode squeezed state (TMSS). Second, the input state and one half of the TMSS are mixed on another beamsplitter and then measured by two homodyne detectors. 
A generalised teleportation circuit is obtained by measuring the general quadratures $\hat x_{\theta_1}$ and $\hat x_{\theta_2}$ while also providing general ancillary states $\ket{\phi_1}$ and $\ket{\phi_2}$ so that
\begin{align}
    \includegraphics{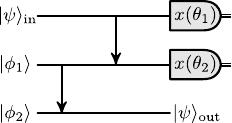}.\label{eq:generalTeleport}
\end{align}
The general mathematical description of \eqref{eq:generalTeleport} is given by the Kraus operator \cite{PhysRevA.102.062411, walshe_streamlined_2021},
\begin{align}
    \ket{\psi}_\text{out}\propto \hat A\big(\ket{\phi_1}, \ket{\phi_2}\big)\hat D(\mu)\hat V\left(\theta_1, \theta_2\right)\ket{\psi}_\text{in},\label{eq:KrausOperator}
\end{align}
where the gate,
\begin{align}
    \hat V(\theta_1, \theta_2)=\hat R\left(-\theta_1\right)\hat P\left(\frac{2}{\tan(\theta_2-\theta_1)}\right)\hat R\left(-\theta_1\right),
    \label{eq:dual_rail_gate}
\end{align}
is a Gaussian unitary dependent on the measured homodyne angles $\theta_1$ and $\theta_2$, the displacement,
\begin{align}
    \mu=-i\frac{m_1e^{-i\theta_2}+m_2e^{-i\theta_1}}{\sin\left(\theta_2-\theta_1\right)},
\end{align}
depends on the measurement results $m_1$ and $m_2$, and the operator,
\begin{align}
    \hat A(\ket{\phi_1}, \ket{\phi_2})=\frac{1}{\pi}\iint d^2\alpha\ \prescript{}{p}{\left<\alpha_I|\phi_1\right>}\cdot\prescript{}{x}{\left<\alpha_R|\phi_2\right>}\hat D(\alpha)
\end{align}
is given by the two ancillary states $\ket{\phi_1}$ and $\ket{\phi_2}$ combined on the first beamsplitter. Here, the subscripts denote the imaginary and real part of $\alpha$, respectively.
Note that the differing relations found in \cite{walshe_streamlined_2021} arise due to the distinct definition of homodyne angles. In the case of ideal CV teleportation with infinitely squeezed states, this leaves,
\begin{align}
    \ket{\psi}_\text{out}\propto \hat A\big(\ket{0}_p, \ket{0}_x\big)\hat D(\mu)\hat V\left(0, \tfrac{\pi}{2}\right)\ket{\psi}_\text{in}\propto \hat D(\mu)\ket{\psi}_\text{in},
\end{align}
highlighting the need for a feed-forward displacement to actually equate in- and output mode. To include the effect of finite squeezing, the following relation will be utilised,
\begin{align}
    \hat N(\beta)\hat x\hat N(-\beta)=\cosh(\beta)\hat x+i\sinh(\beta)\hat p
    \propto \hat S\left(t\right)\hat a\hat S^\dagger\left(t\right),
\end{align}
where $t=\tanh(\beta)^{-\frac{1}{2}}$ and the consequent description of a finitely squeezed state will be,
\begin{align}
    \hat S(t)\ket{\text{vac}}\propto \hat N(\beta)\ket{0}_x,
\end{align}
with $\ket{\text{vac}}$ being the vacuum state.
Since the damping operator commutes with the beamsplitter,
\begin{align}
    \left[\hat B_{12}, \hat N_1(\beta)\hat N_2(\beta)\right]=0,
\end{align}
it can be shown \cite{PhysRevA.102.062411} that its application on the ancillary states $\ket{\phi_1}$ and $\ket{\phi_2}$ of the teleportation circuit can be described by,
\begin{align}
    \hat A\left(\hat N(\beta)\ket{\phi_1}, \hat N(\beta)\ket{\phi_2}\right)=\hat N(\beta)\hat A\left(\ket{\phi_1}, \ket{\phi_2}\right)\hat N(\beta).\label{eq:noiseInKrausOperator}
\end{align}
Hence, the CV teleportation with finitely squeezed states is simply given by,
\begin{align}
    \ket{\psi}_\text{out}\propto \hat N(2\beta)\hat D(\mu)\ket{\psi}_\text{in}.
\end{align}
Next, consider the Knill error correction of the square GKP code, where the squeezed state ancillas are replaced by a specific type of GKP state, the so-called qunaught GKP state,
\begin{align}
    \ket{\varnothing}=\ket{0_{\sqrt{\pi/2}}}_\text{rec},
\end{align}
fulfilling the defining relation,
\begin{align}
    \hat F\ket{\varnothing}=\ket{\varnothing}.
\end{align}
Entangling two qunaught states on a beamsplitter yields an ideal Bell pair of the square GKP code,
\begin{align}
    \hat B_{12}\ket{\varnothing\varnothing}=\frac{1}{\sqrt{2}}\left(\ket{00}_\text{GKP}+\ket{11}_\text{GKP}\right)
\end{align}
effectively turning the CV teleportation circuit into a logical teleportation for GKP qubits.
More precisely, it yields \cite{PhysRevA.102.062411},
\begin{align}
    \hat A\left(\ket{\varnothing}, \ket{\varnothing}\right)&\propto \varnothing\left(\sqrt{2}\hat x\right)\varnothing\left(\sqrt{2}\hat p\right)=\varnothing\left(\sqrt{2}\hat p\right)\varnothing\left(\sqrt{2}\hat x\right)\nonumber\\
    &\propto \hat \Pi_\text{GKP}
\end{align}
where $\varnothing(x) = \prescript{}{x}{\left<x|\varnothing\right>}$ is the position wavefunction and the ideal GKP projector is defined as,
\begin{align}
    \hat \Pi_\text{GKP}=\prescript{}{\text{GKP}}{\ket{0}}\bra{0}_\text{GKP} + \prescript{}{\text{GKP}}{\ket{1}}\bra{1}_\text{GKP}.
\end{align}
Together with Eq.\,\eqref{eq:noiseInKrausOperator}, this results in the output state of a Knill error correction with approximate qunaught GKP states,
\begin{align}
    \ket{\psi}_\text{out}\propto \hat N(\beta)\hat \Pi_\text{GKP}\hat N(\beta)\hat D(\mu)\ket{\psi}_\text{in}.\label{eq:KnillEC}
\end{align}
Remarkably, the application of the ideal GKP projector followed directly by the damping operator ensures that any output state is an approximate GKP state. Consequently, the input state as well as the displacement and first damping operator can only affect the logical content of the state. In other words, the Knill error correction projects any continuous error onto a discrete logical operation within the GKP codespace. The most likely operation to occur depends on the outcome of the homodyne measurements, $\mu$, and requires a correction either in real-time or post-processing. After this logical feed-forward, the likelihood of a remaining logical error decreases with higher quality of the input, $\ket{\psi}_\text{in}$, higher squeezing of the approximate qunaught states, $\beta$, and a smaller deviation from the GKP grid given by the real and imaginary part of $\mu$, namely $\left(\sqrt{2}\mu_R \mod \sqrt{\pi}\right)$ and $\left(\sqrt{2}\mu_I \mod \sqrt{\pi}\right)$.

Another consequence of Eq.\,\eqref{eq:KnillEC} is that the input states do not need to bear any resemblance to GKP states in order to produce a GKP output. Notably, given access to a GKP Bell pair, squeezed states can be used to generate high quality GKP basis states, while the vacuum state \cite{PhysRevLett.123.200502} and other Gaussian states \cite{PRXQuantum.5.020337} are sufficient to generate distillable GKP magic states. Hence, when Gaussian inputs are available, the only non-Gaussian resource needed to perform universal computation with GKP qubits is a supply of approximate qunaught states. 

\subsection{Continuous-variable Macronode Cluster States}

Cluster states are generally considered as a resource in the MBQC model. Here, the cluster state is first generated by entangling a large set of identical qubits, after which computation is performed through sequential single-qubit measurements in different bases. Choosing a specific basis for each qubit then allows one to perform a desired computation. A practical approach to generating such large-scale cluster states is through temporal encoding, where delay lines are used to extend entangled Bell pairs into separate temporal modes, enabling scalable entanglement generation \cite{Larsen_2019, doi:10.1126/science.aay2645}. On the other hand, cluster states can also be understood within the gate-based quantum computing model.
Here, they are interpreted as a set of interconnected macronodes, each containing multiple qubits. At every time step, the qubits of one macronode will first be entangled and then measured, splitting the computation into smaller steps.
The entanglement between macronodes is achieved by initially generating Bell pairs within the given set of qubits, depicted as thick coloured lines between two circles:
\begin{align}\label{eq:GKPBellPair}
    \includegraphics{Figures/GKP_EPR_No_Rotation_light.pdf}
\end{align}
Within a macronode, the entanglement and subsequent measurement of the halves of different Bell pairs teleports the encoded information onto the next macronode whilst applying a specific operation dependent on the chosen measurement bases.

\subsubsection{Dual-Rail Lattice Cluster State}

\begin{figure}[h]
	\includegraphics{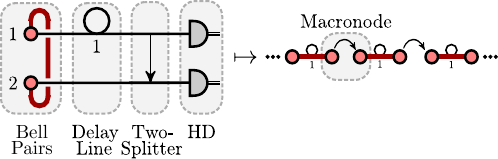}
	\caption{Setup and depiction of the Dual-Rail Lattice cluster state. Each time step, a Bell pair is generated and partially delayed by one clock cycle. The non-delayed half from the current together with the delayed half from the previous time step are then entangled by a beamsplitter, generating a one-dimensional macronode lattice, followed by measurement using two homodyne detectors (HD).}
	\label{fig:DRL}
\end{figure}
The Dual-Rail Lattice (DRL) cluster state is the simplest example of a macronode cluster state. Its GKP Bell pairs are connected by macronodes along a single dimension, resulting in a repeated application of the described Knill error correction. In order to save resources and allow for an ongoing computation, the macronodes are typically separated in time rather than in space. Therefore, GKP Bell pairs are generated at a given clock rate, followed by a time delay of one clock cycle acting on one half of each generated pair. After this redistribution of the entangled qubits across different time steps, a macronode measurement is performed at each clock cycle. Here, the two qubits arriving at a given time are entangled by a beamsplitter and then measured by two homodyne detectors. This setup and the resulting DRL cluster state are shown in Fig.\,\ref{fig:DRL}.

Following Eq.\,\eqref{eq:KrausOperator}, the most general gate that can be performed in one teleportation step is given by Eq.\,\eqref{eq:dual_rail_gate}, with the two homodyne angles $\theta_1$ and $\theta_2$. Using the angles listed in Table~\ref{tab:DRL},
\begin{table}[h]
    \centering
    \caption{Selected measurement bases of the two homodyne detectors of a DRL and the resulting gates.}\label{tab:DRL}
    \begin{ruledtabular}
        \begin{tabular}{CCC}
            \theta_1, \theta_2 & \hat V\left(\theta_1, \theta_2\right) & \text{Logical Gate} \\ \hline \rule{0mm}{\normalbaselineskip}
            0, \frac{\pi}{2} & \hat I & \bar I \\
            -\frac{\pi}{4}, \frac{\pi}{4} & \hat F & \bar H \\
            0, -\arctan(2) & \hat P\left(-1\right) & \bar P
        \end{tabular}
    \end{ruledtabular}
\end{table}
all generators of the single-qubit Clifford gate set can therefore be applied within a single macronode \cite{walshe_streamlined_2021}.

\subsubsection{Quad-Rail Lattice Cluster State}

\begin{figure}
	\includegraphics{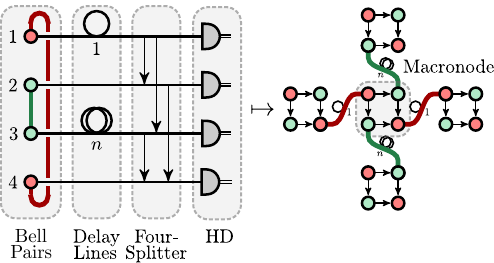}
	\caption{Setup and depiction of the Quad-Rail Lattice cluster state. Each time step, two Bell pairs are generated and partially delayed by one and $n$ clock cycles, respectively. The non-delayed halves from the current together with the delayed halves from previous time steps are then entangled by a beamsplitter network known as foursplitter, generating a two-dimensional macronode lattice, followed by measurements using four homodyne detectors (HD).}
	\label{fig:QRL}
\end{figure}
In order to perform computations with multiple qubits, a two-dimensional cluster state is required. Whilst there are different macronode designs to choose from, the Quad-Rail Lattice (QRL) cluster state has been shown to be favourable due to low gate noise and high flexibility \cite{walshe_equivalent_2023, PRXQuantum.2.030325}. 
It is constructed by equipping two GKP Bell pairs with time delays of one and $n$ clock cycles, respectively, and connecting them by four beamsplitters to create a so-called foursplitter. This setup can be seen in Fig.\,\ref{fig:QRL}.

The macronodes of the QRL form a two-dimensional lattice in time with nearest neighbors connected by GKP Bell pairs. Note that the first dimension, spanned by delays of one clock cycle, exhibits a skewed periodic boundary as $n$ teleportations along it equal one teleportation along the second dimension with delays of $n$ clock cycles. The extent of this first dimension before repeating can therefore be chosen by adjusting the length of the second delay line. When performing a calculation with a static number of $k$ qubits, the choice of $n=k$ would be natural.

\begin{figure}
    \centering
    \begin{subfigure}[t]{\columnwidth}
        \centering
        \includegraphics[]{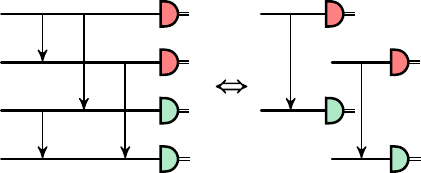}
        \caption{The DRL beamsplitter layer gets removed.}
    \end{subfigure}
    \begin{subfigure}[t]{\columnwidth}
        \centering
        \includegraphics[]{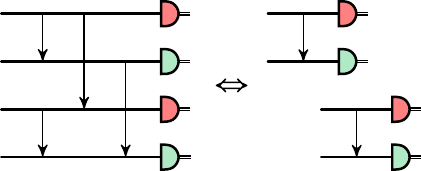}
        \caption{The QRL beamsplitter layer gets removed.}
    \end{subfigure}
    \caption{Reduction of a QRL macronode into two separate DRL macronodes by applying the same measurement bases across modes. Equally colored detectors measure in the identical basis.}
    \label{fig:QRLdecomp}
\end{figure}

While the structure of a macronode cluster state is given by its distribution of the GKP Bell pairs over different time steps, the gates that can be performed are dependent on the beamsplitter network. The foursplitter of the QRL lends its high flexibility and low gate noise from two important symmetries. First, the two beamsplitter layers, further referred to as DRL and QRL layer, commute,
\begin{align}
\includegraphics[]{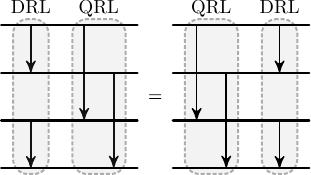}
\end{align}
This means that each of the two layers can be reduced to an addition of measurement outcomes by choosing identical measurement bases for pairs of modes and using,
\begin{align}\label{eq:cancelbs}
\includegraphics[]{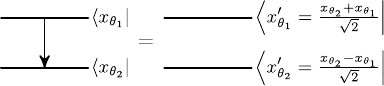}
\end{align}
Note that this reduction is also true for squeezing noise. Consequently, the QRL can be used to perform single-mode teleportations along its different axes with the same amount of squeezing noise as the DRL. The draw-back is that matching the measurement bases means that both single-mode operations performed within the same macronode need to be identical. The reduction of the QRL to different DRLs can be seen in Fig.\,\ref{fig:QRLdecomp}.

Second, any permutation of the input modes commutes with the foursplitter up to permutations and rotations of $\pi$. The specific transformation of the permutation generators has been worked out in \cite{alexander_flexible_2016}. Therefore, any 
 gate that can be performed on a specific combination of in- and output modes can also be performed on any other combination. For example, the two-mode swap gate can be considered as two single-mode identity gates with permuted output states, resulting in a simple permutation of the required measurement bases. A general two-mode gate $\hat{V}_2\left(\theta_1, \theta_2, \theta_3, \theta_4\right)$ -- here acting on modes 1 and 3 -- is given by \cite{walshe_streamlined_2021},
\begin{align}\label{eq:gateV2}
\includegraphics[]{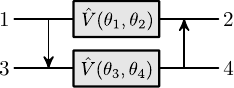}
\end{align}
Using Eq.\,\eqref{eq:CZGateDecomposition}, this leads to the angles required to perform the missing GKP Clifford gate, namely the logical CZ gate $\bar C_Z$. Thus, all GKP Clifford gates can be performed within one teleportation. The angles are listed in Table~\ref{tab:QRL}.
\begin{table}[t]
    \centering
    \caption{Selected measurement bases of the four homodyne detectors of a QRL and the resulting gates.}\label{tab:QRL}
    \begin{ruledtabular}
        \begin{tabular}{CCC}
            \theta_1, \theta_2, \theta_3, \theta_4 & \hat V_2\left(\theta_1, \theta_2,  \theta_3, \theta_4\right) & \text{Logical Gate} \\ \hline \rule{0mm}{\normalbaselineskip}
            0, \frac{\pi}{2}, 0, \frac{\pi}{2} & \hat I \otimes \hat I  & \bar I \otimes \bar I \\
            -\frac{\pi}{4}, \frac{\pi}{4}, -\frac{\pi}{4}, \frac{\pi}{4} & \hat F \otimes \hat F & \bar H \otimes \bar H \\
            0, -\atan(2),  0, -\atan(2) & \hat P\left(-1\right) \otimes \hat P\left(-1\right) & \bar P \otimes \bar P\\
            \frac{\pi}{2}, 0, 0, \frac{\pi}{2} & \text{SWAP}  & \overline{\text{SWAP}} \\
            0, -\atan(2),  0, \atan(2)  & \hat C_Z(1)  & \bar C_Z
        \end{tabular}
    \end{ruledtabular}
\end{table}
It was recently demonstrated, that other two-dimensional cluster state designs can use the concept of the foursplitter to achieve the same noise properties as the QRL \cite{walshe_equivalent_2023}.

\subsection{Topological Quantum Error Correction}\label{sec:TopologicalQEC}

The finite squeezing of GKP states introduces logical errors when performing error correction. Running a multi-qubit computation on the QRL cluster state supplied with approximate qunaught states will therefore lead to logical errors throughout the calculation. Hence, additional qubit error correction on the logical level is needed. As macronodes interact only with their nearest neighbors, interactions between logical qubits are equally limited unless some recurrent rearranging is performed. Consequently, topological quantum error correction codes are a natural option, as they are designed to require only local stabiliser measurements for error correction \cite{bombin2013introductiontopologicalquantumcodes}. Introducing ancillary qubits to perform these stabiliser measurements, topological codes can then be run on qubit lattices that allow only nearest neighbor interactions as well as qubit measurements \cite{fowler}.
Notably, the dimensionality of the underlying qubit lattice induces constraints on the properties of a given topological code. While at least two dimensions are needed for error correction, three dimensions allow codes with transversal universal gate sets \cite{PhysRevLett.102.110502, Bombín_2015} as well as single-shot error correction \cite{Kubica2022}, and four dimensions allow for a property known as self-correction \cite{alicki2008, 10.1063/1.1499754}.

The most prominent topological quantum error correction code is the two-dimensional surface code \cite{fowler}. It requires next neighbor interactions of qubits placed on a two-dimensional square grid. Due to its high fault-tolerance threshold of around 1\% \cite{PhysRevLett.98.190504, fowler}, its simple two-dimensional layout, along with well-researched gates and decoders \cite{PRXQuantum.4.040334, Vuillot2019} it is the common first choice for qubit error correction. In \cite{Litinski_2019} it has been demonstrated how any multi-qubit computation can be performed using patches of surface code in a two-dimensional plane and choosing which stabilisers to measure appropriately. In order to perform non-Clifford gates, magic state distillation \cite{Litinski_2019,Litinski2019magicstate} is required, which takes several physical magic states and grows them into a high quality logical magic state. The necessary real-time feed-forward operations can be reduced to changing the stabilizer measurements of an ancilla patch and can in principle be delayed arbitrarily \cite{Litinski_2019}, limiting the hardware requirements for fault-tolerance and universality to flexible stabilizer measurements and the generation of physical magic states.

The most prominent three-dimensional code is the 3D gauge color code offering both a transversal universal gate set and single-shot error correction \cite{Bombín_2015} but exhibiting lower fault-tolerance thresholds than the 2D surface code. Besides, the mixing of surface and color code has been shown to have advantages in applications such as reducing the overhead of magic state distillation \cite{gidney2024magicstatecultivationgrowing}. In general, providing a flexible setup that allows for different error correction codes is desirable.

\section{Octo-Rail Lattice Configurations}

Given a number of known gate implementations for a particular arrangement of in- and output modes, the same gates can be applied on another mode combination whenever it is the result of a rearrangement of modes given by a permutation in $P_\text{allowed} \subset S_8$. This subgroup $P_\text{allowed}$ consists of all permutations acting on the eight modes going into the eightsplitter that can be compensated for by changing the measurement angles. It can be generated by four double transpositions
\begin{equation}
	P_\text{allowed} = \left<\Big\{\underbrace{(12)(56)}_{P_1},\underbrace{(13)(57)}_{P_2},\underbrace{(14)(58)}_{P_3},\underbrace{(17)(28)}_{P_4} \Big\}\right>,
\end{equation}
where the transposition $(jk)$ swaps modes $j$ and $k$. Alternatively, all pairs of transpositions picked from the following seven sets also generate $P_\text{allowed}$
\begin{align}\label{eq:TranspositionSets}
    &\left\{(12),(34),(56),(78)\right\},\qquad
    \left\{(13),(24),(57),(68)\right\},\nonumber\\
    &\left\{(14),(23),(58),(67)\right\},\qquad
    \left\{(15),(26),(37),(48)\right\},\nonumber\\
    &\left\{(16),(25),(38),(47)\right\},\qquad
    \left\{(17),(28),(35),(46)\right\},\nonumber\\
    &\left\{(18),(27),(36),(45)\right\},
\end{align}
providing an easier way to decide if a given permutation is allowed.
\begin{figure*}[t]
\centering
\includegraphics[width=0.85\textwidth]{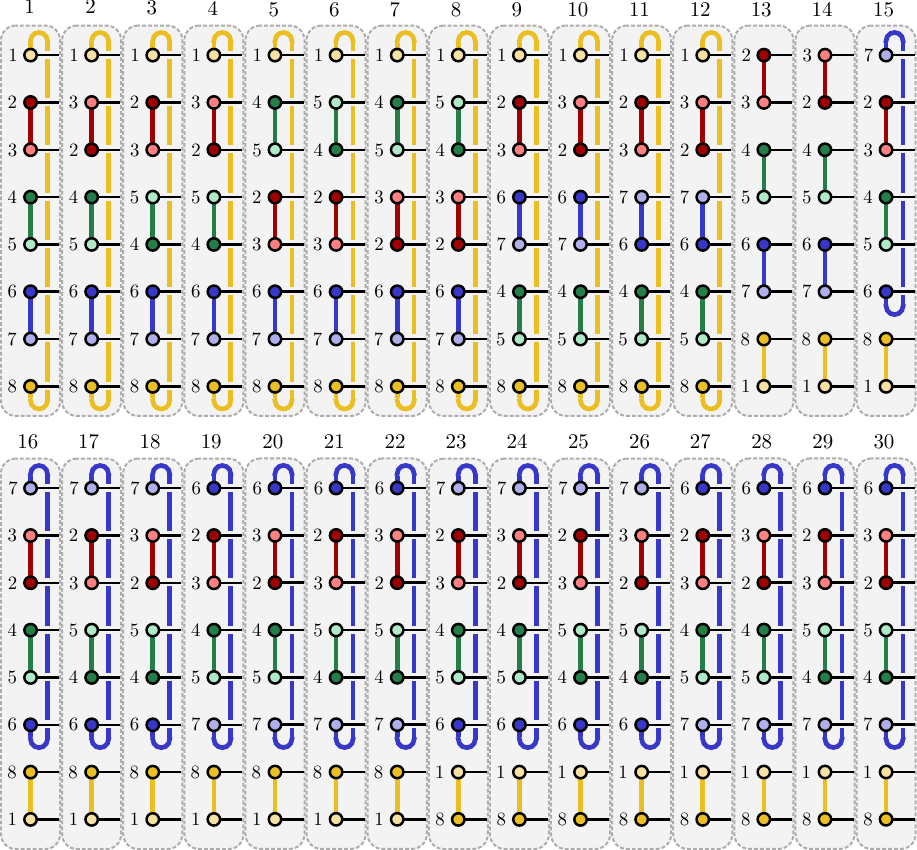}
\caption{Visualisation of the 30 distinct Octo-Rail Lattice configurations, each corresponding to a right coset of $S_8$ split by $P_\text{allowed}$. Each configuration enables the same set of gates to be applied to different combinations of modes.}
\label{fig:input_combination}
\end{figure*}
To give an example, the permutation, 
\begin{align}
	P_\text{test} &= (12)(34)(56)(78),
\end{align}
which can be written as a product of the four elements of the generating set
\begin{align}\begin{split}
	P_\text{test} &= \underbrace{(13)(57)}_{P_2}\underbrace{(14)(58)}_{P_3}\underbrace{(13)(57)}_{P_2}\underbrace{(12)(56)}_{P_1}\\
	&= \underbrace{(13)(14)(13)}_{(34)}(12)\underbrace{(57)(58)(57)}_{(78)}(56)\\
	&= (12)(34)(56)(78),
\end{split}\end{align}
is clearly a product of two pairs of transpositions taken from the first set of Eq.~\eqref{eq:TranspositionSets} and as such $P_\text{test} \in P_\text{allowed}$. For the change of angles to be independent of the chosen gate, the only possible operations acting on the measurement angles are permutations and single-mode rotations. As the eightsplitter does not mix $x$- and $p$-quadratures, the latter only encompasses $\pi$-rotations.
To determine the required change of measurement bases that implement a given input permutation, the following transformation is applied,
\begin{align}
	\vb*{P}_\text{meas} = \mathbf{S} \vb*{P} \mathbf{S}^T,
\end{align}
where $\vb*{P}_\text{meas}$ represents the symplectic transformation corresponding to the required change of measurement bases, $\vb*{P}$ the symplectic transformation associated with an allowed permutation of in- and output modes, and the matrix,
\begin{equation}
	\mathbf{S} = \frac{1}{2\sqrt{2}}\begin{pmatrix}
		1 & -1 & -1 & 1 & -1 & 1 & 1 & -1\\
		1 & 1 & -1 & -1 & -1 & -1 & 1 & 1\\
		1 & -1 & 1 & -1 & -1 & 1 & -1 & 1\\
		1 & 1 & 1 & 1 & -1 & -1 & -1 & -1\\
		1 & -1 & -1 & 1 & 1 & -1 & -1 & 1\\
		1 & 1 & -1 & -1 & 1 & 1 & -1 & -1\\
		1 & -1 & 1 & -1 & 1 & -1 & 1 & -1\\
		1 & 1 & 1 & 1 & 1 & 1 & 1 & 1
	\end{pmatrix}
	\label{eq:eightsplitter_symplectic},
\end{equation}
denotes the symplectic transformation induced by the eightsplitter. Note that only a quarter of the full symplectic matrix is considered in these calculations, as transformations in the $x$- and $p$-quadratures remain decoupled, making it unnecessary to include the complete transformation.

This procedure allows one to determine when a known gate configuration can be re-used on a different set of modes simply by adjusting the measurement bases. On the other hand, it also shows when this is not dynamically possible. Whenever a desired permutation is not in $P_\text{allowed}$, a new static lattice configuration is required to perform the same set of gates. The subgroup $P_\text{allowed} \subset S_8$ splits $S_8$ into 30 right cosets corresponding to 30 distinct ORL configurations shown in Fig.\,\ref{fig:input_combination}. These represent the complete set of inequivalent setups of the Octo-Rail Lattice regarding its available gate set.

\section{Surface Code Basis}

To implement the surface code within the Octo-Rail Lattice (ORL) cluster state architecture, each macronode is measured in a specific basis that enables the X- and Z-stabiliser measurements. Since these measurements require data qubits to be coupled to ancilla qubits via CNOT and CZ gates, respectively, two distinct measurement bases are used for the data-qubit macronodes in the ORL. The choice of measurement angles in these bases directly determines the coupling between each data qubit and its neighbouring ancilla qubits.

\subsection{Even Data Basis}

For the even data macronodes, the following measurement basis is applied:
\begin{equation}
	(\theta_1, \theta_2, \theta_3, \theta_4, \theta_5, \theta_6, \theta_7, \theta_8) = (0,0,0,\tfrac{\pi}{2}, 0, 0, 0, \tfrac{\pi}{2})
\end{equation}

The resulting quadrature transformations, after compensating for displacement known by measurement outcomes, are:
\begin{align}\begin{split}
	x_{1'}' &= -x_1 + \tfrac{1}{\sqrt{2}}(x_{6'} +x_{7'} + 2p_5 - p_6 - p_7 + 2p_8)\\
	p_{1'}' &= -p_1 + \tfrac{1}{\sqrt{2}}(-x_{2'} - x_{3'} + p_2 + p_3 - 2p_4 + 2p_{1'})\\
	x_{2'}' &= \tfrac{1}{\sqrt{2}}(x_{2'} + p_3)\\
	p_{2'}' &= \sqrt{2}p_{2'} + x_1\\
	x_{3'}' &= \tfrac{1}{\sqrt{2}}(x_{3'} + p_2)\\
	p_{3'}' &= \sqrt{2}p_{3'} + x_1\\
	x_{6'}' &= \tfrac{1}{\sqrt{2}}(x_{6'} + p_7)\\
	p_{6'}' &= p_1 + \tfrac{1}{\sqrt{2}}(x_{2'} + x_{3'} + 2p_{6'} - p_2 -p_3 +2p_4)\\ &=  \sqrt{2}p_{6'} + p_5'\\
	x_{7'}' &= \tfrac{1}{\sqrt{2}}(x_{7'} + p_6)\\
	p_{7'}' &= p_1 + \tfrac{1}{\sqrt{2}}(x_{2'} + x_{3'} + 2x_{7'} - p_2 - p_3 + 2p_4) \\ &= \sqrt{2}p_{7'} + p_5'
\end{split}\end{align}
where primed quadratures are after teleportation, non-primed quadratures (except mode 1) are initial GKP qunaught input states before two-mode entanglement generation, and
\begin{equation}
	p_5' = p_1 + \tfrac{1}{\sqrt{2}}(x_{2'} + x_{3'} - p_2 -p_3 +2p_5)
\end{equation}

In this configuration, the quadrature $x_1$ couples to $x_{6'}'$ and $x_{7'}'$, which are associated with measure-$Z$ qubits, while $p_1$ couples to $x_{2'}'$ and $x_{3'}'$, associated with measure-$X$ qubits. Although $p_{6'}'$ and $p_{7'}'$ involve more terms and thus appear to be more affected by finite squeezing noise than $p_{2'}'$ and $p_{3'}'$, $x_1$ itself includes noise from the second teleportation step through the previous macronode. This ensures that the total noise is balanced between measure-$Z$ and measure-$X$ qubits. This can be seen more clearly by expressing $p_{6'}'$ and $p_{7'}'$ in terms of the intermediate quadrature $p_5'$, making their structure analogous to that of $p_{2'}'$ and $p_{3'}'$, with $p_5'$ taking the role of $x_1$.

The qunaught states used as input are GKP states with quadrature peak spacing of $\sqrt{2\pi}$. Ignoring finite squeezing, quadrature terms scaled by $\sqrt{2}$ correspond to displacements of $2n\sqrt{\pi}$ for integer $n$, which do not affect the logical GKP qubit. Terms scaled by $1/\sqrt{2}$, however, result in displacements that can cause bit and phase flips. These correspond to the expected back-action from stabiliser measurements in the surface code and commute with the stabilisers, preserving the encoded quantum information. In the presence of finite squeezing, these terms contribute Gaussian noise, which is corrected during each teleportation step via standard GKP error correction.

Besides the quadrature transformation shown above, the quadratures are displaced determined by the measurement outcomes of the modes in the macronode ($m_{1,\dots,8}$). These are:
\begin{align}\begin{split}
	x_{1'}' &: \tfrac{1}{\sqrt{2}} (2m_1 + m_2 + m_3 -2m_5 + m_6 +m_7)\\
	p_{1'}' &: \tfrac{1}{\sqrt{2}} (2m_1 - m_2 - m_3 +2m_5 + m_6 +m_7)\\ 
	p_{2'}' &: \tfrac{1}{\sqrt{2}} (- m_3 - m_4 - m_7 - m_8)\\
	p_{3'}' &: \tfrac{1}{\sqrt{2}} (- m_2 - m_4 - m_6 - m_8)\\
	p_{6'}' &: \tfrac{1}{\sqrt{2}} (-2m_1 + m_3 + m_4 -2m_5 - m_7 - m_8)\\
	p_{7'}' &: \tfrac{1}{\sqrt{2}} (-2m_1 + m_2 + m_4 -2m_5 - m_6 - m_8)
\end{split}\end{align}  

\subsection{Odd Data Basis}

For the odd data macronodes, the following measurement basis is applied:
\begin{equation}
	(\theta_1, \theta_2, \theta_3, \theta_4, \theta_5, \theta_6, \theta_7, \theta_8) = (\tfrac{\pi}{2},0,0,0, 0, 0, 0, \tfrac{\pi}{2})
\end{equation}

The resulting quadrature transformations, after compensating for displacement known by measurement outcomes are:
\begin{align}\begin{split}
	x_{1'}' &= -x_1 + \tfrac{1}{\sqrt{2}}(-x_{2'} -x_{3'} + p_2 + p_3 + 2p_5 + 2p_8)\\
	p_{1'}' &= -p_1 + \tfrac{1}{\sqrt{2}}(x_{6'} + x_{7'} - 2p_4 - p_6 - p_7 + 2p_{1'})\\
	x_{2'}' &= \tfrac{1}{\sqrt{2}}(x_{2'} + p_3)\\
	p_{2'}' &= -p_1 + \tfrac{1}{\sqrt2}(2p_{2'} + x_{6'} + x_{7'} - 2p_4 -p_6 -p_7)\\ &= \sqrt{2}p_{2'} + p_5'\\
	x_{3'}' &= \tfrac{1}{\sqrt{2}}(x_{3'} + p_2)\\
	p_{3'}' &= -p_1 + \tfrac{1}{\sqrt2}(2p_{3'} + x_{6'} + x_{7'} - 2p_4 -p_6 -p_7)\\ &= \sqrt{2}p_{2'} + p_5'\\
	x_{6'}' &= \tfrac{1}{\sqrt{2}}(x_{6'} + p_7)\\
	p_{6'}' &= \sqrt2 p_{6'}-x_1\\
	x_{7'}' &= \tfrac{1}{\sqrt{2}}(x_{7'} + p_6)\\
	p_{7'}' &= \sqrt2 p_{7'}-x_1
\end{split}\end{align}
where
\begin{equation}
	p_5' = -p_1 + \tfrac{1}{\sqrt2}(x_{6'} + x_{7'} - 2p_4 -p_6 -p_7)
\end{equation}
This time the $x_1$ quadrature is coupled to $p_{6'}'$ and $p_{7'}'$ and $p_1$ is coupled to $p_{2'}'$ and $p_{3'}'$. Besides this, the same comments can be made as for the even data qubits.

The known displacement are,
\begin{align}\begin{split}
	x_{1'}' &: \tfrac{1}{\sqrt{2}} (2m_1 + m_2 + m_3 + m_6 + m_7 - 2m_8)\\
	p_{1'}' &: \tfrac{1}{\sqrt{2}} (2m_1 - m_2 - m_3 + m_6 + m_7 + 2m_8)\\
	p_{2'}' &: \tfrac{1}{\sqrt{2}} (2m_1- m_3 - m_4 + m_5 + m_6 - 2m_8)\\
	p_{3'}' &: \tfrac{1}{\sqrt{2}} (2m_1- m_2 - m_4 + m_5 + m_7 - 2m_8)\\
	p_{6'}' &: \tfrac{1}{\sqrt{2}} (m_3 + m_4 + m_5 + m_6)\\
	p_{7'}' &: \tfrac{1}{\sqrt{2}} (m_2 + m_4 + m_5 + m_7)
\end{split}\end{align} 

\section{Non-Square GKP Encodings}

For the square GKP encoding, the angles needed to perform different logical gates via the teleported gate
\begin{align}
    \hat V(\theta_1, \theta_2)=\hat R(-\theta_1)\hat P\left(\frac{2}{\tan(\theta_2-\theta_1)}\right)\hat R(-\theta_1)
\end{align}
are known. In order to perform the same logical gates in a non-square GKP encoding given by
\begin{align}
    \ket{j}_\text{GKP$^\prime$}=\hat U_{G}\ket{j}_\text{GKP}=\hat R(\omega_1)\hat P(\lambda)\hat R(\omega_2)\ket{j}_\text{GKP}
\end{align}
one needs to apply the gate
\begin{align}
    \hat V_{\text{GKP}'}\left(\theta_1, \theta_2\right)=\hat U_G \hat V\left(\theta_1, \theta_2\right)\hat U_G^\dagger.
\end{align}
This can be achieved by enacting the teleported gate
\begin{align}\begin{split}
    \hat V\left(\phi_1,\phi_2\right)&=\hat U^{\dagger T}_{G}\hat V\left(\theta_1,\theta_2\right)\hat U^\dagger_{G}\\
    &=\left(\hat U_{G}\hat U^T_{G}\right)^\dagger\hat V_{\text{GKP}'}\left(\theta_1, \theta_2\right).
\end{split}\end{align}
To find the transformation of angles from $(\theta_1, \theta_2)$ to $(\phi_1, \phi_2)$ we use the notation
\begin{align}
    \hat V'(\theta, \sigma)=\hat R(\theta)\hat P(\sigma)\hat R(\theta)=\hat V(-\theta, \arctan(\tfrac{2}{\sigma})-\theta)
\end{align}
and solve the equation
\begin{align}
    \hat V'(\phi, \tau)=\hat U^{\dagger T}_{G}\hat V'(\theta,\sigma)\hat U^\dagger_{G}
\end{align}
for $\phi$ and $\tau$. Introducing $\phi'=\phi+\omega_1-\tfrac{\pi}{2}$ and $\theta'=\theta-\omega_2-\tfrac{\pi}{2}$ this can be written as
\begin{align}\begin{split}\label{eq:Vexpanded}
    \hat R(\phi')\hat P(\tau)\hat R(\phi')&=\hat F^\dagger\hat P(-\lambda)\hat F\hat R(\theta')\hat P(\sigma)\\
    &\times\hat R(\theta')\hat F\hat P(-\lambda)\hat F^\dagger.
\end{split}\end{align}
The LDU and UDL decompositions of its symplectic matrix give the following two decompositions of a rotation
\begin{align}
    \hat R(\alpha)&=\hat P(\tan\alpha)\hat S(\cos\alpha)\hat F\hat P(\tan\alpha)\hat F^\dagger\\
    &=\hat F^\dagger\hat P(\tan\alpha)\hat F\hat S^\dagger(\cos\alpha)\hat P(\tan\alpha)
\end{align}
as long as $\cos\alpha\neq0$.
Consequently, the right-hand side of Eq.~\eqref{eq:Vexpanded} becomes
\begin{widetext}\begin{align}\begin{split}
    &\hat F^\dagger\hat P(\tan\theta'-\lambda)\hat F\hat S^\dagger(\cos\theta')\hat P(\sigma+2\tan\theta')\hat S(\cos\theta')\hat F\hat P(\tan\theta'-\lambda)\hat F^\dagger\\
    =&\hat F^\dagger\hat P(\tan\theta'-\lambda)\hat F\hat P\left(\frac{\sigma+2\tan\theta'}{\cos\theta'}\right)\hat F\hat P(\tan\theta'-\lambda)\hat F^\dagger\\
    =&\hat R(\phi')\hat P^\dagger(\tan\phi')\hat S(\cos\phi')\hat P\left(\frac{\sigma+2\tan\theta'}{\cos\theta'}\right)\hat S^\dagger(\cos\phi')\hat P^\dagger(\tan\phi')\hat R(\phi')\\
    =&\hat R(\phi')\hat P\left(\frac{\cos\phi'}{\cos\theta'}\left(\sigma+2\tan\theta'\right) - 2\tan\phi'\right)\hat R(\phi')
\end{split}\end{align}\end{widetext}
where $\phi'=\arctan(\tan\theta'-\lambda)$ was fixed along the way. Thus, we find
\begin{align}
    \phi&=\arctan(\tan\theta'-\lambda) + \tfrac{\pi}{2} - \omega_1\\
    \tau&=\frac{\cos\phi'}{\cos\theta'}\left(\sigma+2\tan\theta'\right) - 2\tan\phi'
\end{align}
and finally
\begin{align}\begin{split}
    \phi_1&=\omega_1-\tfrac{\pi}{2}-\arctan(\cot(\theta_1+\omega_2)-\lambda)\\
    \phi_2&=\arctan\bigg[1\Big/\bigg(\frac{\sin(\phi_1-\omega_1)}{\sin(\theta_1+\omega_2)}\big(\cot(\theta_2-\theta_1)\\
    &+\cot(\theta_1+\omega_2)\big)-\cot(\phi_1-\omega_1)\bigg)\bigg] + \phi_1.
\end{split}\end{align}
In case that $\cos\theta'=0$ we quickly obtain
\begin{align}
    \phi_1&=\omega_1\\
    \phi_2&=\arctan\left(\frac{1}{\cot(\theta_2-\theta_1)-\lambda}\right) + \phi_1. 
\end{align}

\clearpage

\end{document}